# Learning from metastable grain boundaries

**Avanish Mishra[1], Sumit A. Suresh[2], Saryu J. Fensin[2], Nithin Mathew[1*], and Edward M. Kober[1*]**

[1]Theoretical Division (T-1), Los Alamos National Laboratory, Los Alamos, 87545, NM, USA

[2]Materials Physics and Applications (MPA-CINT), Los Alamos National Laboratory, Los Alamos, 87545, NM, USA

*mathewni@lanl.gov (Nithin Mathew) and emk@lanl.gov (Edward M. Kober)

Grain boundaries (GBs) govern critical properties of polycrystals. Although significant advancements have been made in characterizing minimum energy GBs, real GBs are seldom found in such states, making it challenging to establish structure-property relationships. This diversity of atomic arrangements in metastable states motivates using data-driven methods to establish these relationships. In this study, we utilize a vast atomistic database (~5000) of minimum energy and metastable states of symmetric tilt copper GBs, combined with physically-motivated local atomic environment (LAE) descriptors (Strain Functional Descriptors, SFDs) to predict GB properties. Our regression models exhibit robust predictive capabilities using only 19 descriptors, generalizing to atomic environments in nanocrystals. A significant highlight of our work is integration of an unsupervised method with SFDs to elucidate LAEs at GBs and their role in determining properties. Our research underscores the role of a physics-based representation of LAEs and efficacy of data-driven methods in establishing GB structure-property relationships.

## Introduction

The differently oriented crystals or grains in polycrystalline materials are separated by interfaces or boundary regions known as grain boundaries (GBs)[1-3]. GBs are classified according to the geometrical relationships between the adjoining grains[4,5]. These distinct GBs affect the physical[6-8], chemical[9,10], and mechanical response[11-16] a of material differently. For example, the shorter bond length due to boundary reconstruction in Aluminum (Al) Σ9 GB structure, increases its tensile strength compared to single-crystal (sc) Al[13]. The reduction in phonon mean free path due to the presence of GBs also decreases the thermal conductivity of material. However, other than the distribution of GB, the reduction in phonon mean free path also depends on the GB structure, i.e., the local arrangement of atoms at the GB[17]. Similarly, GB structure also influences various other properties, such as electrical conductivity/resistivity[18-20], corrosion[9,21-23], ductility[4,16,24], strength[25,26], etc.[27-30]. Such correlations between GB structure and properties pioneered the emergence of GB engineering[31-33], where the type and distribution of GBs are tailored to achieve the desired material response. Nonetheless, most of the computational studies to date have focused on well-defined GBs with equilibrium and minimum energy structures[34-38]. Real GBs seldom have such well-defined structures, and therefore, it is vital to study structure-property relationships over a wide range of GB structures.

Among the commonly known properties, grain boundary energy (GBE) and energy density (atomic energy normalized by volume associated with an atom) are directly linked to the atomic arrangement (local atomic environments or LAEs) at the boundary[39-42], which makes them primary candidates for elucidating structure-property relationships. After one considers all possible macroscopic (five GB characters) and microscopic degrees of freedom, a large number of GB structures (metastable states) can be generated. The extent of possible GB structures presents a significant challenge for the utilization of data-driven methods to learn broad correlations and predict structure-property relationships. Earlier, such attempts to predict/optimize GB energy using data-driven methods were limited to minimum energy structures of special GBs[43,44]. The structural features used in these studies included structural unit model[31,43,44], GB dislocation arrays[45-49], common neighbor analysis[50], polyhedral template matching[51], and polyhedral unit model[52]. While these descriptors can characterize and estimate a property of interest for known or ordered structural arrangements at GB, these do not work well for metastable GB structures. In addition, structure-property relationships derived from these LAEs are also insufficient in describing the properties of GBs in polycrystalline materials due to the presence of junctions, amorphous-like GBs, and diverse distribution of GBs[36,53]. Therefore, a more robust and exact description of LAEs is needed to develop accurate predictive models for properties of arbitrary GBs.

A variety of atomic environment descriptors, such as Smooth Overlap of Atomic Positions (SOAP)[54,55], bi-spectrum coefficients[56,57], moment tensor representation[58], etc., have been emerging from the field of machine-learning (ML) derived interatomic-potentials. SOAP, which uses radial and spherical harmonic bases to construct atomic environment descriptors, has been used with ML to understand structure-property relationships for GB properties. For example, Rosenbrock et al.[39] used SOAP to identify structural building blocks at GBs and predicted average quantities such as GBE, GB mobility, and shear coupling. Wagih et al.[41] used SOAP to predict segregation energy for polycrystalline materials at different fidelity. In another work, SOAP is utilized to predict atomic energy spectra for GB atoms[40]. More recently, Homer et al.[59] predicted the GBE for ~7000 GBs of Al using SOAP descriptors. Similarly, other above-mentioned interatomic potential descriptors are also used to predict various properties of materials. Nevertheless, the inherent challenge associated with these descriptors are computational cost, limited physical insight, and lack of completeness/redundancy, which limits the prospect of establishing a generalized structure-property relationship. For example, the above-mentioned studies have utilized 1000-3000 SOAP descriptors to predict different properties, which makes it challenging to derive physical insights[17,39,41]. In their recent study, Song et al.[40] attempted to reduce >1000 SOAP vectors to a few for predicting GB energy density for a smaller dataset. However, a dependence on deviation from the perfect lattice structure is the only physical insight derived in their study. The redundancy in these descriptors also reduces the interpretability of the derived ML models.

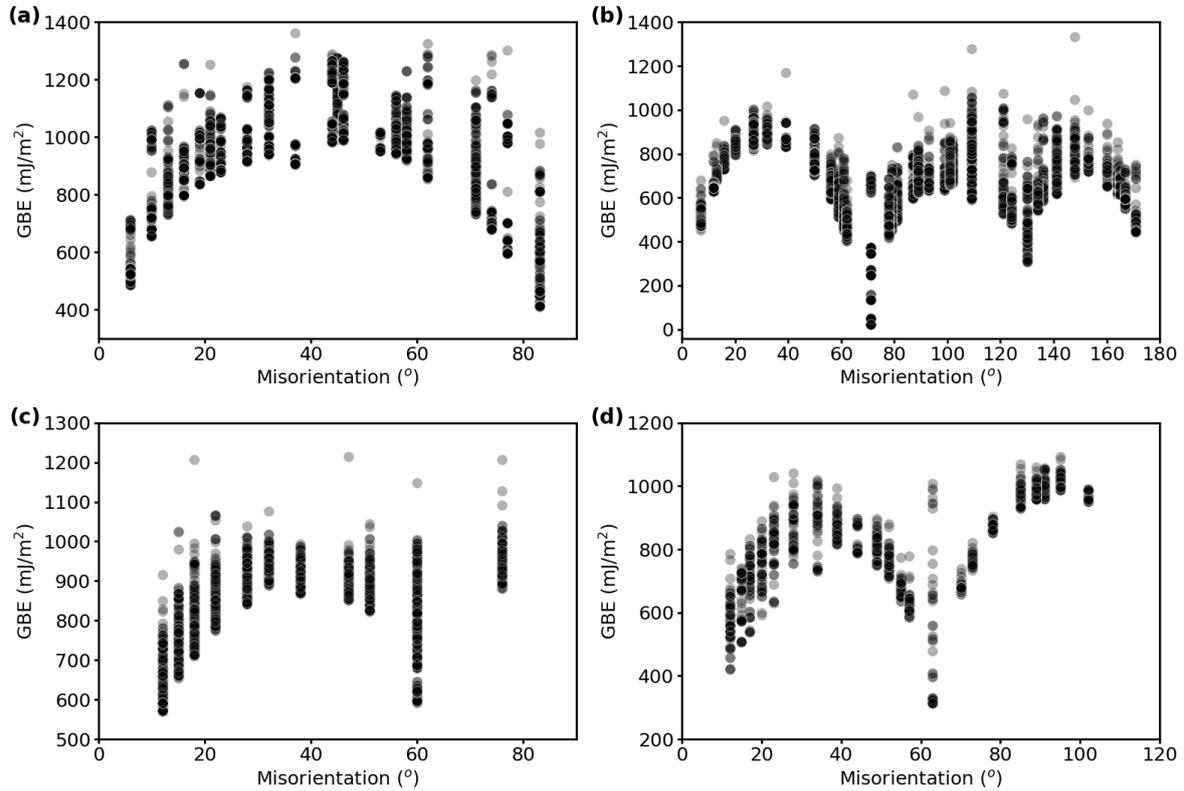

*Figure 1, Grain boundary energy (GBE) at different misorientation angles for (a) <100>, (b) <110>, (c) <111>, and (d) <112> minimum-energy and metastable symmetric tilt boundaries of FCC Cu. Scatter points have reduced transparency and are represented by a pale grey color, where the near coincidence of multiple points results in the darker regions.*

To overcome these challenges, we have utilized Strain Functional Descriptors (SFDs)[60], a convergent and symmetry-adapted set of descriptors to represent the LAEs in GB structures. The desired structure-property relationship is established by training ML models on a vast database of four different symmetric tilt grain boundaries (<100>, <110>, <111>, and <112>) of Cu comprising of over 5000 minimum energy and metastable GB structures, as shown in Figure 1. The observed feature importance from regression models explains the role of density and deformation moments of SFDs in predicting GB properties, thus providing physical insights. Furthermore, physically meaningful atomic environments at GBs are identified using an unsupervised Gaussian mixture model (GMM)[61]. A total of four different sets of 'GB features' are derived from 19 per-atom SFD descriptors (based on shape and size). Tree-based regression models are developed using these features to predict GBE and energy density with unprecedented accuracy. Although the model trained using mean SFD features results in superior performance for GBE prediction, ML models based on GMM probability, frequency, and cosine similarity helps to establish a correlation between various modes of deformation at GB with physical properties. Despite being trained only on symmetric tilt GBs, the diversity of metastable structures results in robust predictive capability for the ML models, which is verified by predicting atomic energy density for new (unseen) nanocrystalline microstructures. This establishes the generality of our database and the ability of our models to predict properties of arbitrary/general GBs. Our study reveals the importance of a physics-informed representation of the LAEs using SFDs to establish the various GB structure-property relationships. The superiority of SFDs combined with the wide variety of atomic environments in the large database of metastable GB structures, opens up an exceptional prospect for developing structure-property relationships for real GBs.

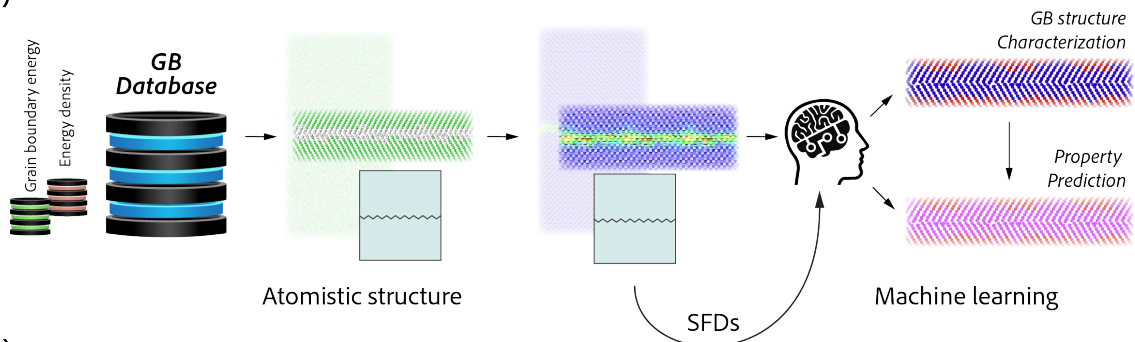

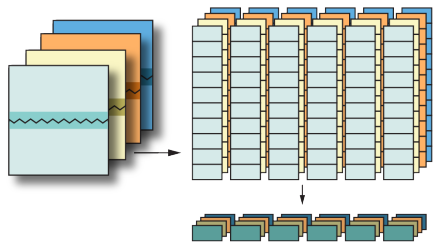
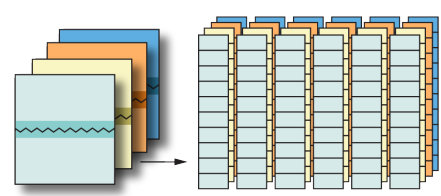
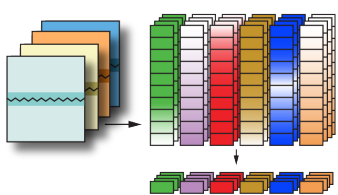
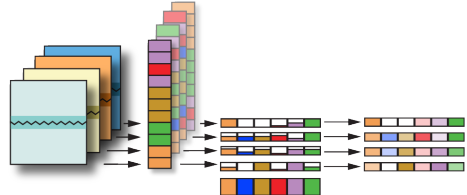
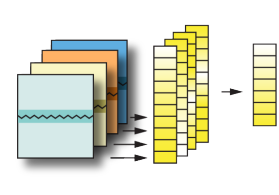

*Figure 2, (a) Schematic showing framework utilized for characterizing GB structure and developing ML models. (b) Highlights different feature-sets derived from SFDs to develop ML models.*

## Results

**Grain boundary energy (GBE) prediction**
**Regression models using statistical features of SFDs**
We first considered GBE for four symmetric tilt boundaries as a property of interest to develop ML models, using a tree-based random forest regression. The workflow is depicted in Figure 2(a). The initial set of ML models are trained using mean SFDs of all atoms in the 'GB region' (as shown in Figure S1) as features, Figure 2(b)-i. From each data set, 10% is considered as unseen data for testing and the ML model is trained on 80% of the remaining data, which is randomly selected, with 20% of it is used for validation (Figure 3). The performance of the regression model is checked by measuring the coefficient of determination ($R^2$) and root-mean-squared error (rmse). We also verified the robustness of the ML model by training it on 1000 different train-validation splits and 5-fold cross validation repeated 3 times to avoid any bias in the data selection process; the related error histogram and cross validation score are shown in Figure S2. The best regression model for each symmetric tilt is selected. The training $R^2$ for all cases is greater than 0.99, whereas the validation $R^2$ value is greater than 0.94 with exceptionally low rmse values, as shown in Figure 3. The $R^2$ value for these models trained separately on specific symmetric tilts exceeded 0.91 for test (unseen) data, indicating a high level of fit between the measured and predicted values.

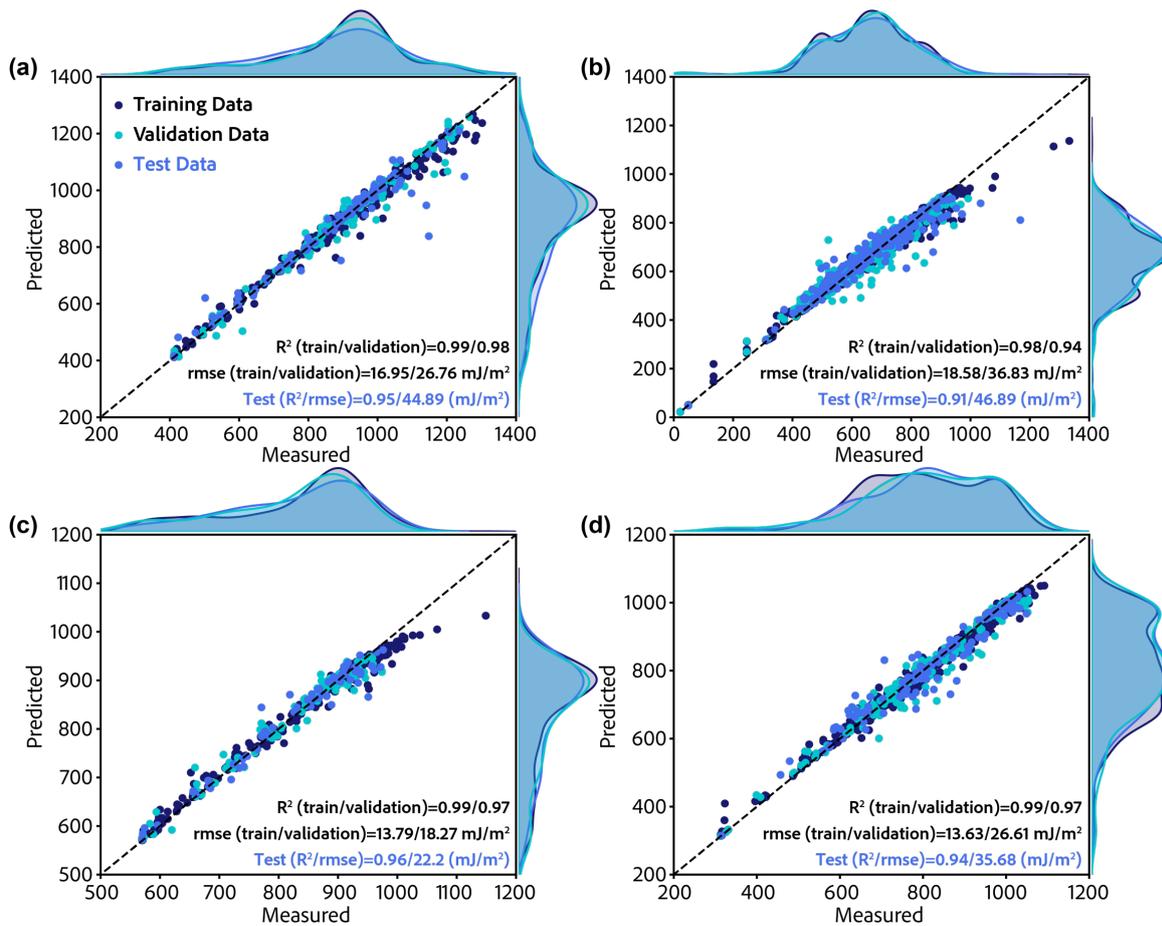

*Figure 3, Scatter plot showing measured vs. predicted values for (a) <100>, (b) <110>, (c) <111>, and (d) <112> symmetric tilt GBs. Training data is shown by the dark blue dots and are in the back layer, with the Validation data being turquoise dots in the next layer, and the Test data is shown by blue dots in the top layer. Here, test is performed on 10 % of data that the model has not seen during training.*

By extracting the feature importance from these models, we can derive physical insights into the effect of dominant deformation modes on GB energy. As detailed in Kober et al.[60], SFDs can be mapped to size (i.e., density), shape, and orientation metrics. The 19 SFD vectors used in our study consists only of size and shape metrics. This choice highlights the influence of density, nature of deformation (strain, strain gradients etc.), and directionality of deformation (tetragonal, orthorhombic, deviations from tetrahedral etc.) terms on predicting GBE. Figure 4 shows the feature importance from the random forest regression models for each symmetric tilt dataset. For the <100> and <111> symmetric tilts, shown in Figures 4(a) and 4(c), respectively, the most important feature is P4_I8, which represents a density term indicative of volumetric deformation. This term correlates well with the excess volume at the GB, which is known to correlate well with GBE[53]. The next most important term is P4_I6, which is a metric for the type of shear the neighborhood is undergoing. (It is closely related to P2_I0, which also measures the net shear, but has a different radial weighting.) Next in importance are the third order deformation terms P3_I0 and P3_I1, which are strain gradient terms. These summarize the net third-order deformation and the deviation from tetrahedral symmetry, respectively. The <111> GB also depends strongly on P4_I2, which is a metric for the distortion away from octahedral/FCC symmetry. We also trained the regression model using only the three most important features for <100> and <111> tilt GBs. For both cases, developed models show a reduction in performance; nonetheless, prediction accuracy remains high, as shown in Figure S3.

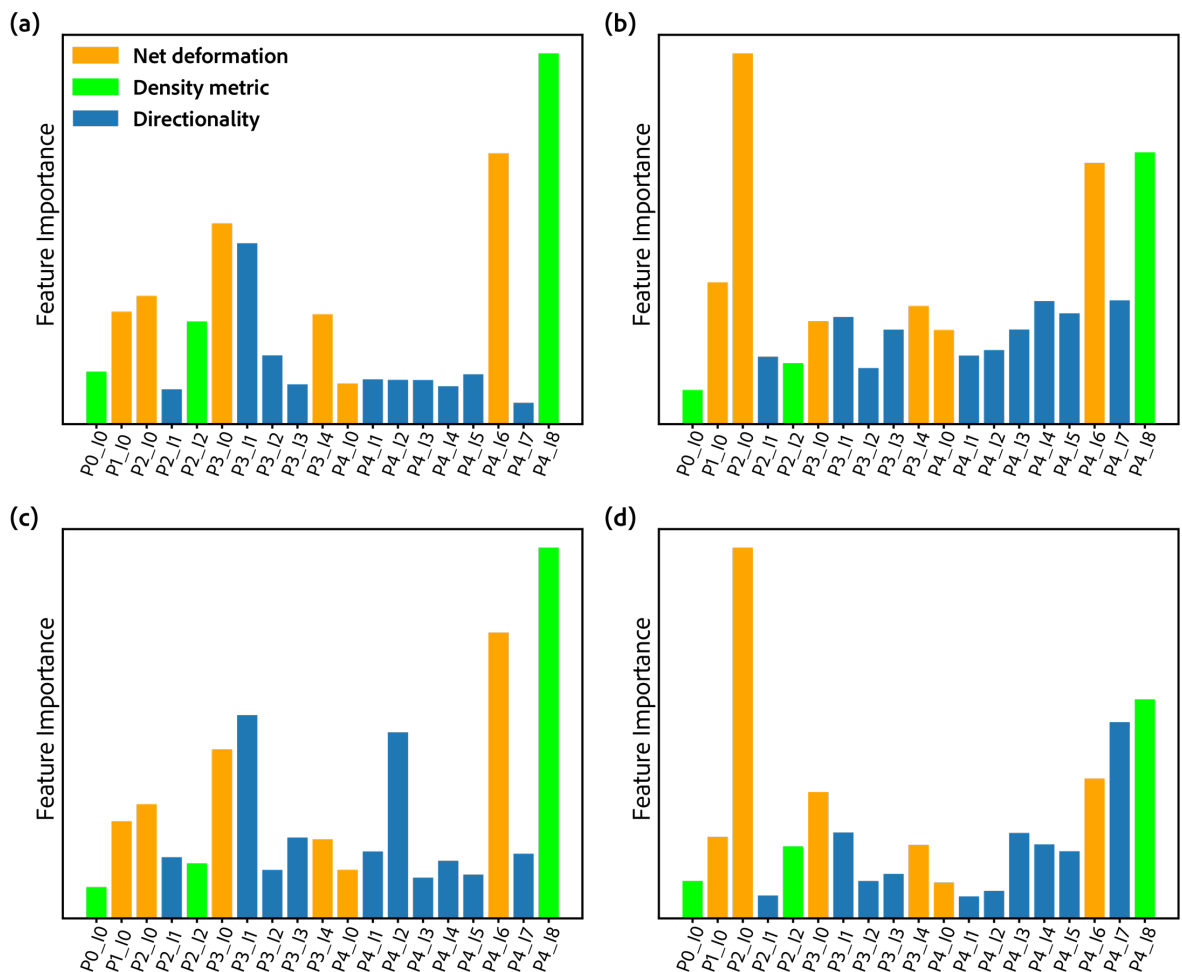

*Figure 4, Feature importance from individual models developed for (a) <100>, (b) <110>, (c) <111>, and (d) <112> symmetric tilt GBs. The value of feature importance is shown by the height of the bars. Net deformation, density metrics, and directionality terms of SFDs are shown by orange, green, and blue bars, respectively.*

For the <110> and <112> symmetric tilts, shown in Figures 4(b) and 4(d), respectively, the most significant feature is P2_I0, with P4_I8 following in importance. As mentioned above, P4_I8 correlates with the excess volume at GB, while P2_I0 is net shear/deviatoric strain. As described in the methods section, the protocol for generating the database consists of in-plane translations of the crystals relative to each other. Given that <110> is a close-packed direction in FCC, multiple sheared states are accessible during in-plane translations in along the tilt axis, and this results in relative higher importance of P2_I0. The P4_I6 term is also an important feature which is closely related to P2_I0, as mentioned above, having a different radial weighting. P4_I6 is very important for <111> and is reasonably significant for <112> GB. The P4_I7 term is moderately important for the <112> GB, where this is a metric for the character of the shear (tetragonal vs. orthorhombic). Akin to <100> and <111> tilts, we also trained ML models using the three most important features for <110> and <112> symmetric tilts. These ML models also show a minor reduction in performance, though the prediction accuracy is still outstanding, as shown in Figure S3. Moreover, the observed similarity in feature importance for <100> and <111> or <110> and <112> symmetric tilts also illustrate the higher fidelity of SFDs. For example, <110> and <112> symmetric tilts may exhibit HCP stacking faults separating partial dislocations at GB and therefore, they are expected to have similar deformation features.

The bivariate distribution plot of GBE with the average value of SFDs in GB region also demonstrates the observed feature importance. This is illustrated in Figure 5 for the <112> symmetric tilt GBs. For example, a linear correlation of important features (P2_I0 and P4_I8) with GBE can be seen in Figure 5. Interestingly as P2_I0 is the most important feature according to Figure 4(d), it has the most linear relationship or high correlation with GBE. Fairly good linear correlations with the other significant features for the <112> boundaries (P4_I6, P4_I7 and P3_I0) are also apparent. Bivariate distributions for other tilt boundaries are provided in the supporting information (Figure S4). In every case, the observed feature importance is clearly illustrated in bivariate plots. It is interesting to note that in addition to the density (P0_I0, P2_I2, and P4_I8), gradient in density (P1_I0), and deviatoric strain (P2_I0) metrics, a strong correlation of GBE is predicted with higher order deformation terms such as P3_I0-P3_I4 (functions of strain gradients) and P4_I0, P4_I6, P4_I7 (functions of $2^{nd}$ derivatives or curvatures of the strain). For the first time, this clearly demonstrates the importance of these higher order deformation terms in predicting GBE.

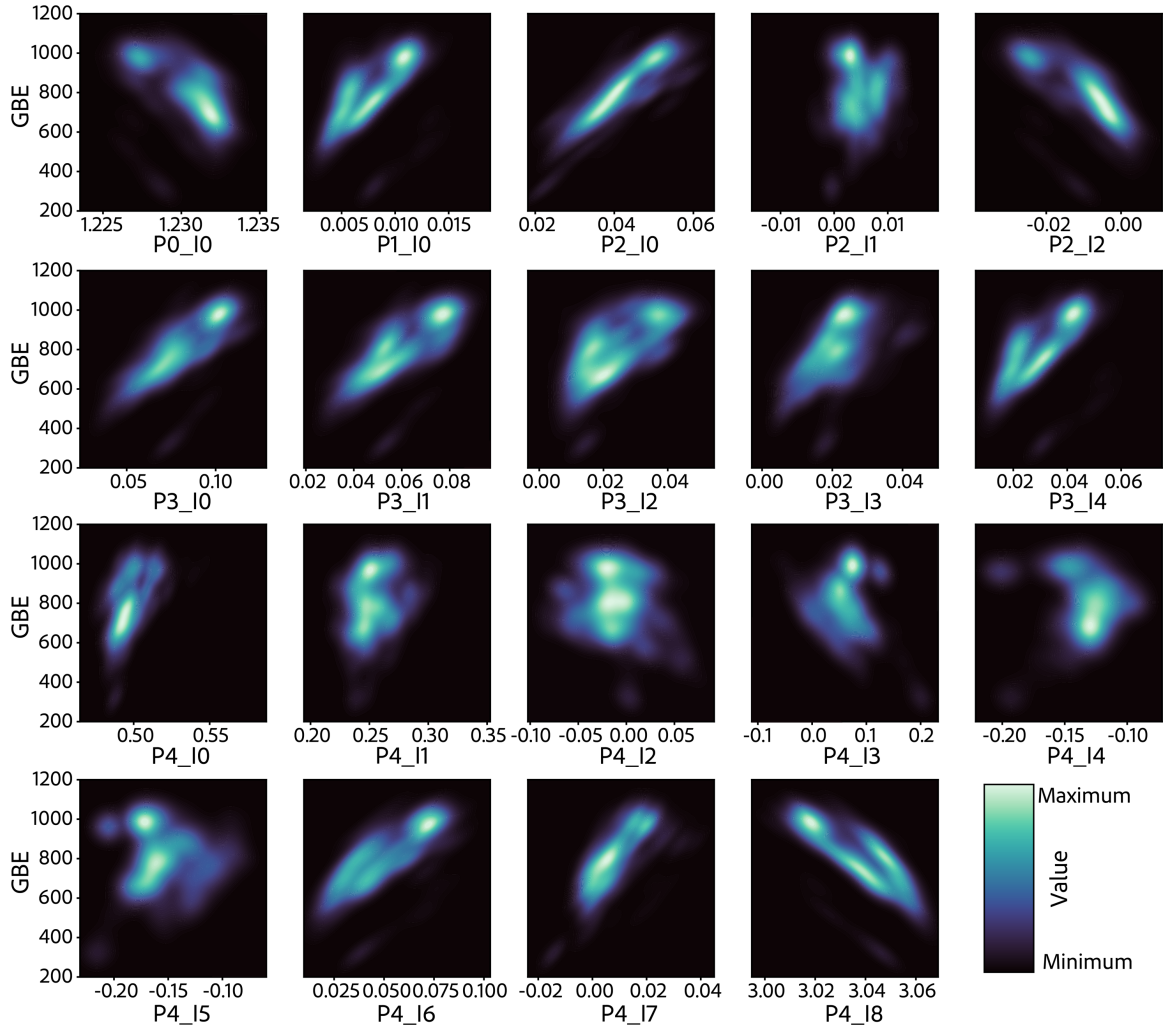

*Figure 5, Bivariate distributions of grain boundary energy (GBE) and mean SFD features using kernel density estimation for <112> symmetric tilt boundaries. Features of higher importance show linear variation with GBE.*

Furthermore, to examine the robustness of SFDs, we also developed regression ML models using another statistical feature, the kurtosis of SFD vectors. The kurtosis measures the width of the distribution of the functions, and this emphasizes the significance of the tail of the distribution or involvement of outliers in interpreting the data. Similar to mean SFDs, each GB is represented by a kurtosis SFD vector of 19 components. Next, we trained ML models for predicting GBE using kurtosis SFD vectors. The performance of ML models shows a slight reduction in the test data prediction for all four symmetric tilt boundaries, as shown in Figure S5. Nonetheless, these models still show exceptional performance, such as in the case of <112> symmetric tilt GBs, the train $R^2$ is similar to the model using mean SFDs (0.99), whereas validation $R^2$ is marginally reduced to 0.96 from 0.97. A similar change in the performance for other symmetric tilts is also observed. The observed feature importance is different from the previous model which uses mean SFD features, as can be seen from Figure S5. Though, it will show some common important features that exhibit a more significant deviation from a normal distribution. Overall, the ML model developed using statistical SFD features predicts GBE with unprecedented accuracy.

*May 31, 2024*

**Regression models using features derived from Gaussian Mixture Model (GMM)**

Next, we wanted to develop ML models to predict GBE using features from the second approach that relies on the population of the GMM classes. Based on clustering tests on minimum energy <100> symmetric tilt GBs using GMMs, in conjunction with Davies-Bouldin[62] and Calinski-Harabasz[63] scores, we developed six-class GMM models to divide the GB atoms into different atomic environments. Each GMM class represents a certain type of GB atom defined in terms of SFDs. Due to the physical interpretability of SFDs, any ML model based on GMM features will help us understand the role of these atomic environments in predicting GBE. To examine different LAEs at GB, the mean value of SFDs for different GMM classes is used to assign atoms to classes, as shown in Figure 6(a) for the representative case of <112> symmetric tilt GBs. Further, to facilitate data analysis, the values are color-coded to indicate whether they are the minimum or maximum for that descriptor. Several examples of <112> GBs, which are color-coded by their classification, are shown in Figures 6(b)-(f). The distribution of GMM classes effectively captures the atomic arrangement at GBs. Specifically, the core of the GB with a 12° misorientation angle (Figure 6(b)) primarily consists of GMM classes #1 and #0. In contrast, the surrounding area, where atoms were removed during the GB structure generation, has classes #2, #5, and #3. For a GB with a 70° misorientation angle shown in Figure 6(c), where no atoms were deleted, the distribution of GMM classes across the GB remains relatively uniform, with the core predominantly occupied by GMM class #3. This stark contrast in atomic deletion is notably evident in the analysis of minimum energy and metastable GBs that share the same GB plane ([1,3,1]) and a misorientation angle of 63°, as depicted in Figures 6(d) and (e). The primary difference between these GBs is the deletion of an atom within the red circle in Figure 6(e). Comparing the same regions within both GBs (marked by a black dotted circle), the impact of this deletion is observed as a variation in GMM classification of the metastable GB structure compared to the minimum energy case. This highlights the capability of GMM classes, identified using SFDs, to detect even minor modifications at GBs.

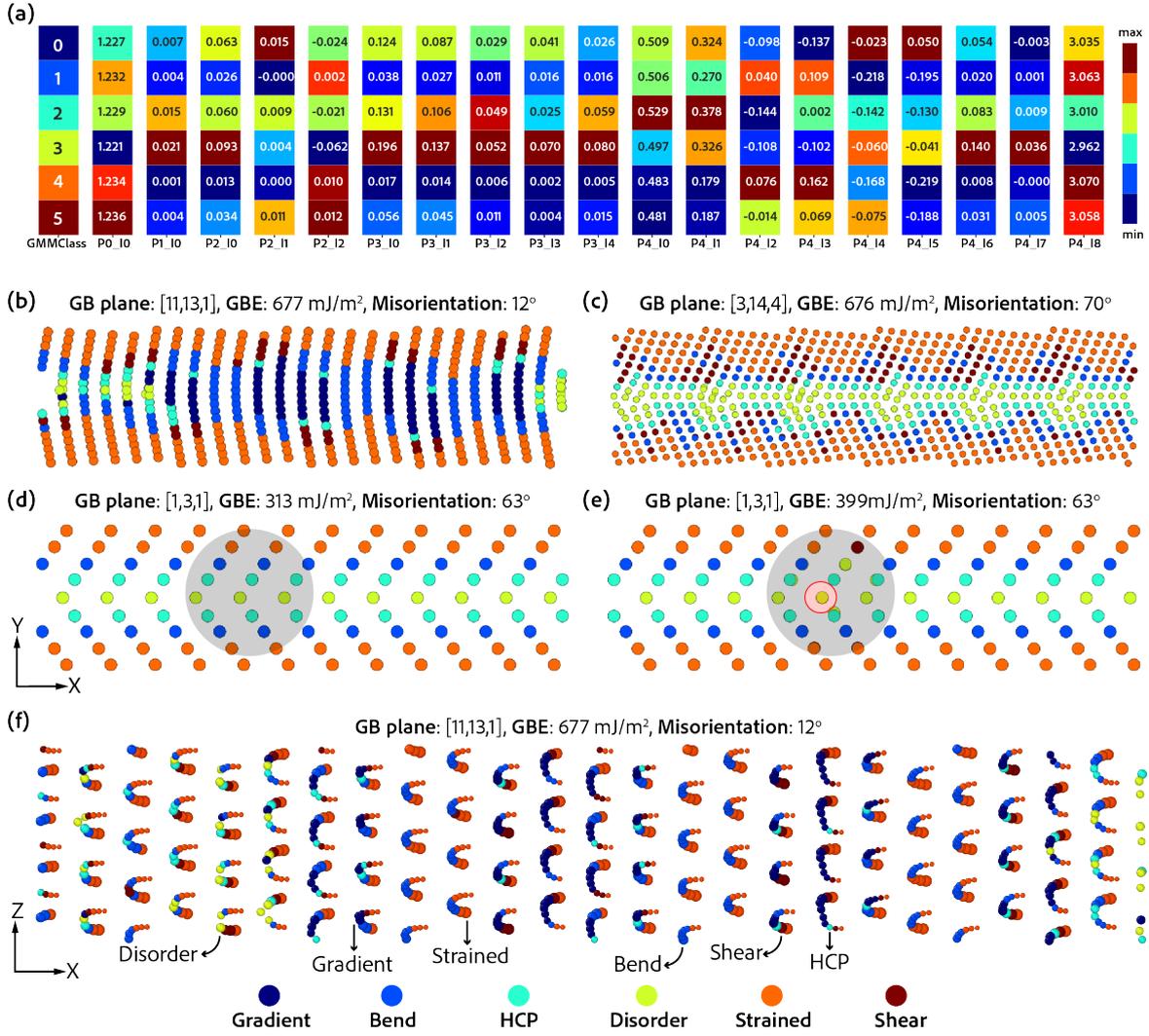

Figure 6, (a) Mean value of SFDs for each GMM class for <112> symmetric tilt. Each column is colored using a "jet" color map, blue-to-red representing minimum-to-maximum values. (b-e) shows example structures of <112> symmetric tilt boundaries with GMM class (atom type) distribution; color-coded for GMM classes. GB structures (d) and (e) only differ by the deletion of one atom in (e), highlighted by a red circle. The black circle highlights the change in GMM class distribution in (e) around the deleted atom, whereas the same circle in (d) displays the ideal case. (f) top view for the GB region in (b); here, atom sizes vary with their position along the Y direction (in the plane of view) to highlight the atomic arrangement in the GB region. The atomic arrangement in (f) agrees with the physical interpretation of these classes from SFDs.

The class that stands out the most is class #4 (orange), which has the minimum SFD values (represented by the dark blue boxes in the row) and is referred to as "Strained" FCC. In an ideal scenario, this class would be FCC (face-centered cubic) if it had no density gradient (P1_I0 = 0), no net shear (P2_I0 = 0), and no net strain gradient (P3_I0 = 0 and P3_I4 = 0). It's important to note that since we have excluded most FCC atoms from this six-class GMM by initially creating a two-class GMM model to separate the bulk and GB atoms, the mean SFD values are not zero. From Figure 6, it is readily apparent that these connect the GB to the bulk region. The next two classes with a considerable number of low values (blue boxes) are #1 (blue) and #5 (dark red). These are more distorted FCC structures. Although similar, class #5 has a larger value of net shear (P2_I0), while class #1 has a larger value of curvature (P3_I3). Upon visual inspection of Figures 6(b) and 6(f), it is evident that class #1 is clearly associated with regions of bending of the lattice within a low-angle tilt boundary. Similar inspections show that class #5 is undergoing a shear distortion, particularly for the planar sliding motion that distorts towards an HCP geometry. Classes #5 and #1 will be labeled as "Shear" and "Bend", respectively, for simplicity. Next, we examine the class #2 atoms (turquoise), which have substantially different values compared to the FCC class, particularly large shear and strain gradient values. Upon visual inspection, it is obvious that these atoms have structures similar to "HCP", although they are often heavily sheared, particularly towards the FCC geometry. Their proximity to the "Shear" class (class #5) highlights the FCC-to-HCP transformation associated with the GB strain fields. Class #0 also significantly differs from the FCC class. It exhibits moderate gradients in density (P1_I0), net shear (P2_I0), and net strain gradient (P3_I0, P3_I4), leading to rather complex deformations. Visual inspections suggest a gradual deformation of the lattice, and this class will be labeled as "Gradient". Finally, class #3 has the most substantial differences from the FCC class, and these typically form the core of the GB. Visual inspection reveals a variety of highly distorted structures, and this class will be designated as "Disorder".

Next, ML models are developed for <112> symmetric tilt GBs using average GMM class probability (Figure 2(b)-iii), and frequency of GMM classes (Figure 2(b)-iv). They show almost identical prediction performance, as shown in Figures 7(a) and 7(b). These ML models do not outperform the model developed using mean SFDs as features, however. For example, the validation $R^2$ for GMM feature models for <112> symmetric tilt is 0.94, lower than 0.97 of the mean SFD model, as shown in Figure 7. Similar performance is also observed for other tilt boundaries, as shown in Figures S6-S8. The feature importance for model using GMM class probability is shown in Figure 7(c). In the case of <112> tilt GBs, the probability of class #3 is the most important feature, followed by classes #1 and #5 highlighting the importance of the "Disorder", "Bend", and "Shear" atoms in predicting GBE. The Strained class (#4) has the highest frequency as compared to other GMM classes (Figure 7(d)) but does not have high feature importance, demonstrating that the strain FCC atoms are relatively less important in determining the GBE, which is intuitive.

In the case of <100>, <110>, and <111> symmetric tilts regression models, developed using GMM class probability and frequency of GMM classes as features, classes #2, #3, and #4 are most important, as shown in Figures S6-S8 respectively. Note that GMM models are developed separately for each symmetric tilt, resulting in varying GMM class numbering for each case; thus, representing the different LAEs. In contrast, the dominant classes for these tilts are GMM classes #4, #1, and #0, respectively. For <100>, the important atom type of class #2 is located at the core of GB; in <110>, class #3 constructs the GB region, and class #4 in <111> tilt is observed away from the core of the GB. Consequently, ML models developed using GMM-based features help us understand the correlation between physically meaningful atomic environments and GBE. For every tilt boundary, it explains how the distribution of the atom type (broadly atomic arrangement) governs the GB property. Similar behavior was predicted for models developed with the frequency of GMM classes as features. In addition, our observations suggest that the importance of a specific atom type or GMM class may vary across different symmetric tilt boundaries. We also developed a regression model with cosine similarity metrics of GB atoms as features (Figure 2(b)-v) to predict GBE, as shown in Figure S9 and the details are discussed in Supplementary Information section S4.

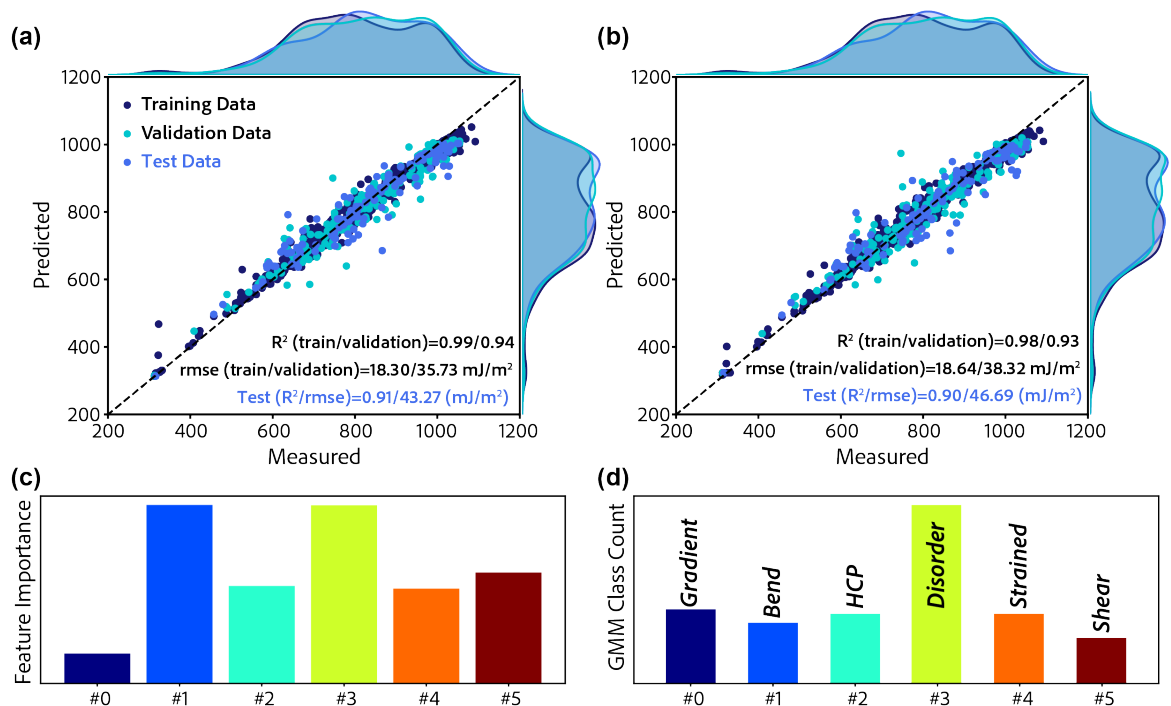

*Figure 7, Scatter plot showing measured vs. predicted values for the model developed using (a) average GMM class probability and (b) frequency of GMM classes as features for <112> symmetric tilt. (c) Feature importance and (d) count of different GMM classes in <112> symmetric tilt boundaries.*

**Combined regression model for all symmetric tilt GBs**

Subsequently, we develop a generalized (combined) regression model for predicting GBE of different tilt boundaries. As mean SFD features exhibited the best performance for individual tilt boundaries, we used them to develop ML models described in this section. In the first attempt, we used a model trained on <112> boundaries to predict GBE for <100>, <110>, and <111> symmetric tilts, as shown in Figure 8(a). The model fails to predict GBE for other boundaries due to the distinctive distribution of SFDs for different tilt boundaries (Figure S10). Therefore, we created a mixed dataset from four tilt GB types consisting of 5479 structures to develop a combined model. 80 % of the combined dataset (4382 GBs) is used for model training, whereas the rest, 20 % (1097), is treated as unseen data for testing. The train and validation $R^2$ for the random-forest regression model on the combined dataset are 0.99 and 0.93, respectively (Figure 8(b)). Figure 8(d) shows that density (P4_I8) and deformation (P2_I0 and P4_I6) terms are important features. Our ML model for GBE with 19 SFDs shows superior predictive performance ($R^2$ of 97%) compared to a previously reported model, developed from a database of 388 CSL GBs using over 3000 SOAP descriptors, with a performance of 89%[39]. An expansive recent study, which spanned all the five macroscopic degrees of freedom and used over 1000 SOAP descriptors reported a predictive performance of 94%[59]. The comparison with this model is not straightforward as the databases, although of similar size, span different macroscopic degrees of freedom. The performance of our combined SFD model on the different symmetric tilt datasets are shown in Figure 8(c). The $R^2$/rmse values for different symmetric tilt datasets are mentioned as color legends in Figure 8(c). The combined model for predicting GBE would likely be further improved by including more data from other types of boundaries (mixed, twist, asymmetric, etc.) and interfaces.

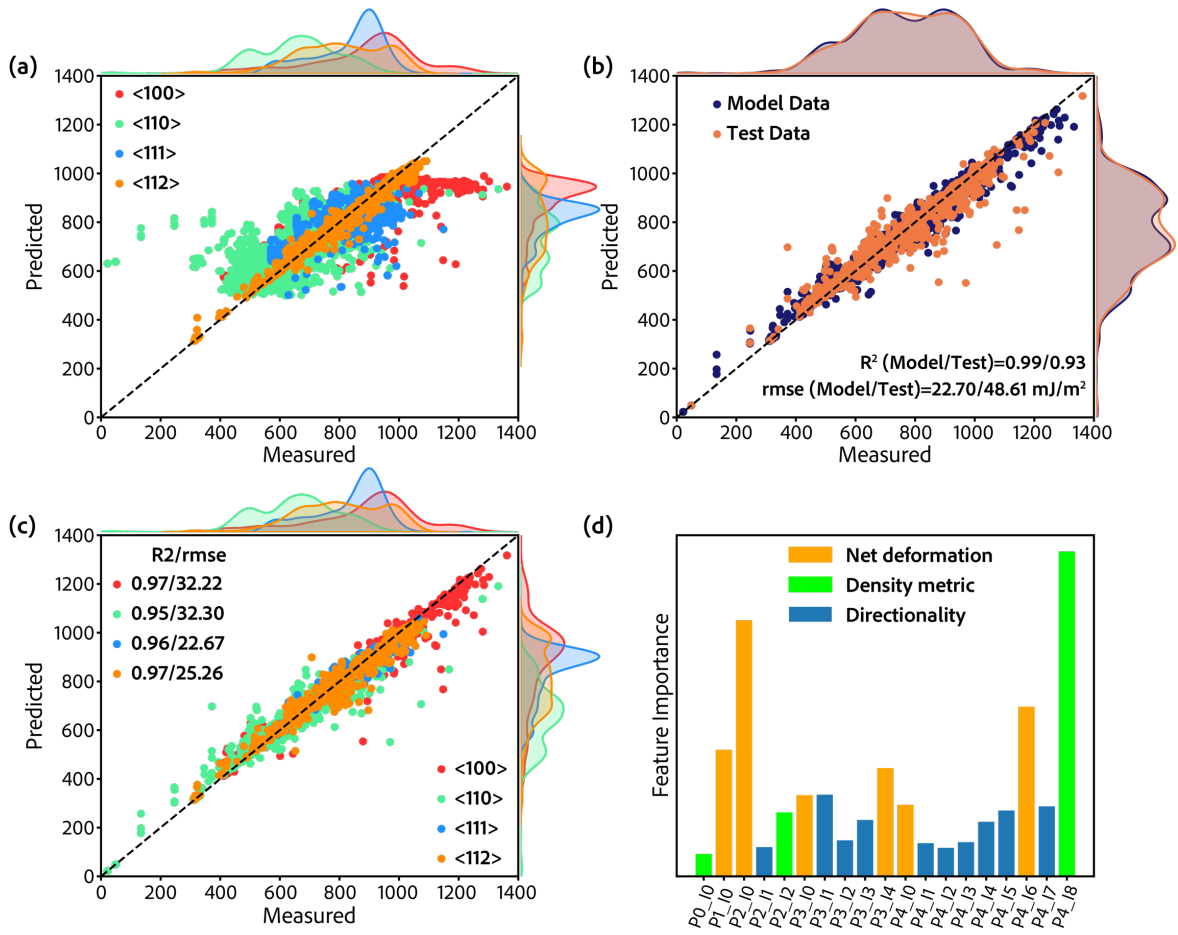

*Figure 8, (a) Scatter plot showing measured vs. predicted GBE for all symmetric tilt boundaries by the model developed using only <112> tilt data as the training dataset. Scatter plot showing measured vs predicted GBE for (b) combined model and (c) performance of combined model on individual GB sets. Model data in (b) consists of combined training and validation data. The color labels in (c) are $R^2$/rmse values for respective GB sets as defined in (a). (d) Feature importance from the combined model. Net deformation, density metrics, and directionality terms of SFDs are shown by orange, green, and blue bars.*

**Energy density**

**Energy density predication using per-atom SFDs**

In this section, we ascertain the predictive capability of our ML models to general GBs. Expanding our scope, we consider predicting properties for GBs in nanocrystalline materials with varied GB distributions. As it is challenging to define an area-normalized GBE in nanocrystalline samples, we develop models to predict atomic energy density, as outlined in methods. Since energy density is a per-atom property, the first challenge in developing these models is addressing the size of the dataset. When considering only the GB region, the dataset for each symmetric tilt is comprised of ~1-3 million data points (atoms). Thus, training an ML model with these many data points would be a daunting task. Therefore, we developed some techniques to curate our database. We started by randomly selecting 12 thousand (12K) data points for training/validating the model, and the rest was used as test data. The model developed using 12K data points for <112> symmetric tilt GBs shows excellent performance with train/ validation $R^2$ of 0.999/0.988 (*rmse*: 0.00034/0.00108 eV/Å³). Nonetheless, as shown in Figure S11, the model does not accurately predict the higher atomic energy density due to the skewness in the distribution of energy density values. For example, the distribution of $E_{den}$ for <112> symmetric tilt shows maximum frequency in the range of -0.32 to -0.25 eV/Å³, whereas the higher energy density region has almost no data points (see Figure 9(c)).

To develop a more general model, we considered rmse of data sets of various sizes for <112> symmetric tilt GBs as shown in Figure 9(a). It can be seen that the error is greatly reduced with randomly selected 200K data points. To further improve the model for test data, we used a different sampling method as shown in Figure 9(b): out of the 200K data points, 180 K are randomly sampled from the whole distribution. For the remaining 20K, we first selected 40K from higher atomic energy density values (non-overlapping from 180K) and randomly sampled 20K from this set. Thus, our training dataset (for train and validation) comprises 200K data points which span the full range of atomic energy density distribution. The performance of these models is shown in Figure 9(c) (randomly selected) and 9(d) (curated). As expected, the curated model shows superior performance on the test data set which consisted of ~843K points ($R^2$ and rmse values at bottom-right corners in Figures 9(c) and (d)). The training performance, predictions on test data, and corresponding features importance for the model trained on randomly selected 200K data points, as well as curated 200K data points for all four symmetric tilts, are shown in Figures S12-S15. In all these cases the density metric, P2_I2, is the most important feature for predicting atomic energy density.

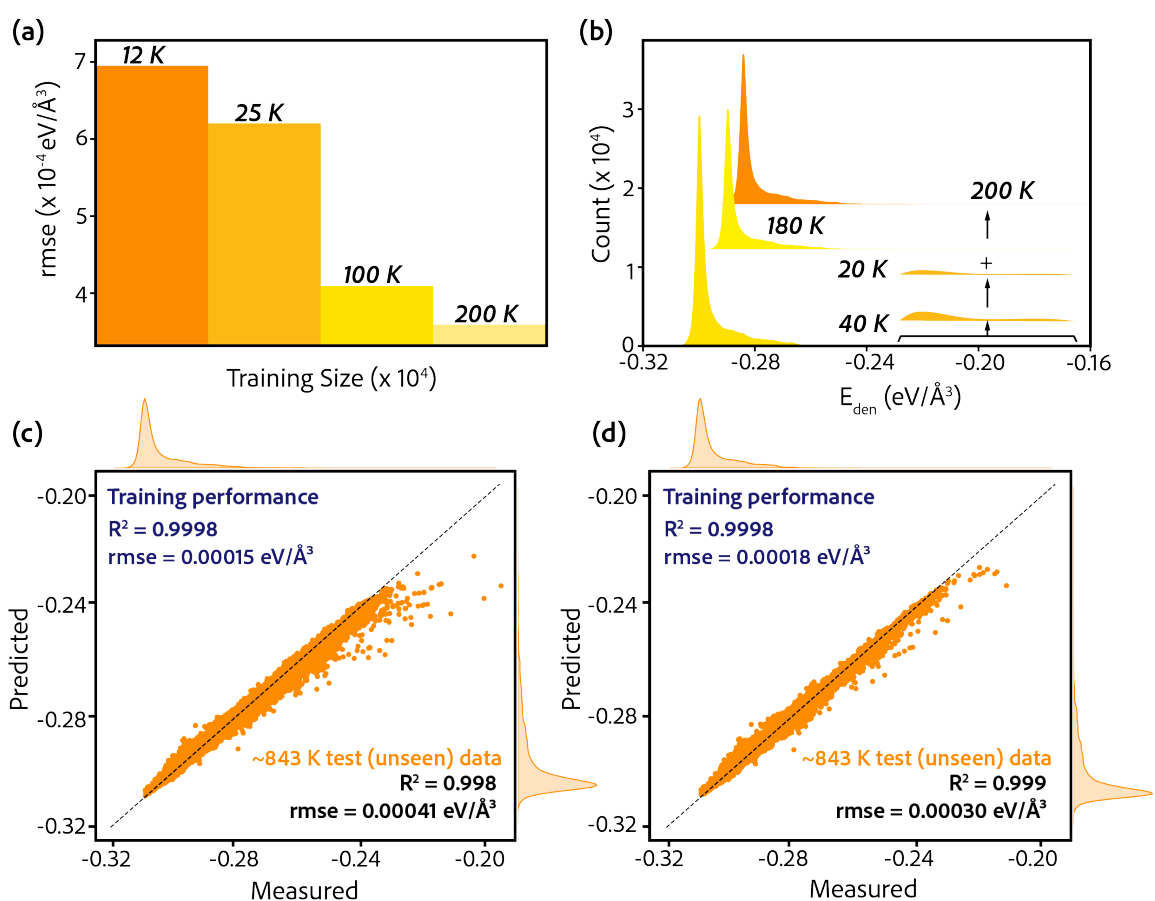

*Figure 9, (a) Validation data prediction error (rmse) for models trained on 12K, 25K, 100K, and 200K atomic energy density data points of <112> symmetric tilt GBs. Here, data points are selected randomly from $E_{den}$ distribution. (b) Fixing the problem with outliers; for a dataset with 200K points, 180K are randomly selected, whereas 40K are selected from the tail of the distribution, out of which 20K are used to construct a 200K dataset. Scatter plot showing measured vs. predicted atomic energy density for a model trained on (c) randomly selected 200K data and (d) 200K data with fixed outliers for <112> symmetric tilt GBs. Feature importance and training performance for these models are shown in Figure*

S15.

Based on these observations, the training dataset for a generalized model of energy density is constructed to span the full range by combining data from metastable states of all symmetric tilts. This is done as following: we select 50K data points each from four symmetric tilt GBs (total of 200K), where we randomly sample 40K from the full range, and 10K from higher energy density values, resulting in a training data set of 200K data points. The developed model exhibits exceptional $R^2$ (rmse) of 0.999/0.999 (0.0002/0.0005 eV/Å$^3$) for train/validation (Figure 10 (a)), whereas model applied to ~6M unknown data points predicts atomic energy density with $R^2$/rmse of 0.998/0.0004 eV/Å$^3$ (Figure 10 (b)). In comparison, the ML model trained on randomly selected 200K of combined dataset shows higher error (rmse=0.0005 eV/Å$^3$) as shown in Figure 10(d). The ability of the generalized model to predict outliers is clearly visible in Figure 10(b) as opposed to Figure 10(d). We also predicted atomic energy density for all the atoms in our bicrystal samples (not only the GB region), as shown in Figure S16. Overall, the model predicted the atomic density for full structure with a maximum error of only ±1%, whereas the error for minimum energy boundaries is negligible (~0.1 %).

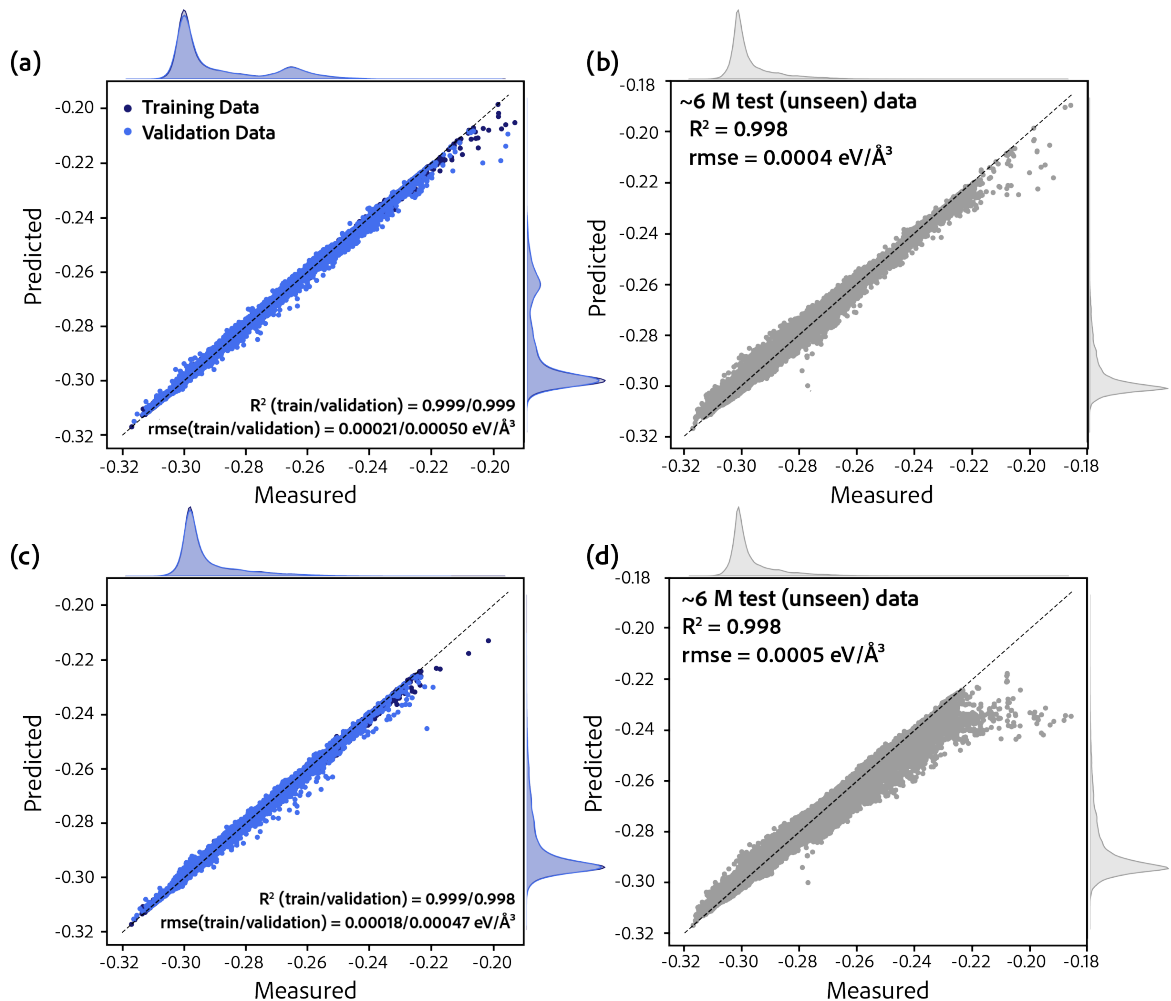

*Figure 10, Scatter plot showing measured vs. predicted atomic energy density for a model trained on a combined dataset of (a) curated 200K data and (c) randomly selected 200K data. Scatter plots in (b) and (d) showing measured vs. predicted atomic energy density for the rest of 6 million atoms using models developed in (a) and (c), respectively.*

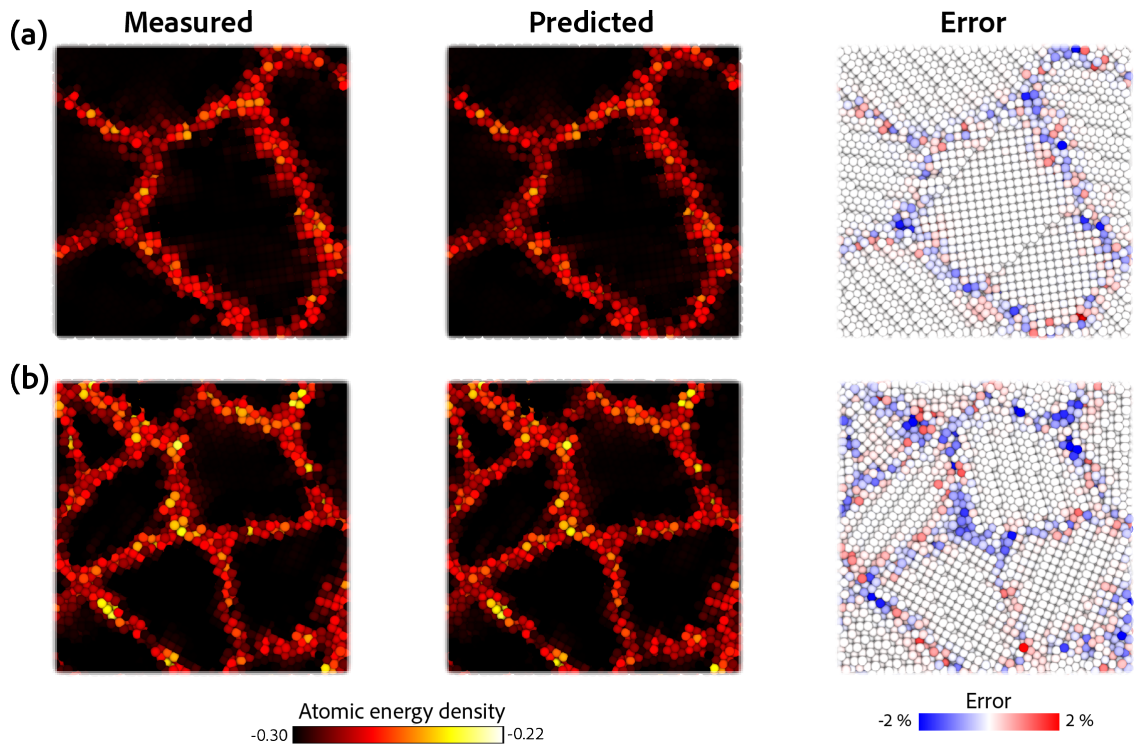

*Figure 11, Actual and predicted atomic energy density and prediction error for nanocrystal of 10 nm$^3$ with (a) 3 and (b) 15 grains. ML model predicts the atomic energy density for these nanocrystals with an error of ±2 %*

   To test the predictive capability of our models for arbitrary GBs, we developed a test database of nanocrystalline microstructures using the same interatomic potential. We built nanocrystals with grain size of ~10 nm$^3$ with randomly oriented grains, with the number of grains ranging from 3 to 25. These nanocrystals have different GBs, triple junctions, and GBs that are not necessarily symmetric tilts but of different characters. Nonetheless, the combined regression model (from Figure 10(a)) predicts atomic energy density for these nanocrystals with exceptional accuracy, an average error of ±2 % (Figure 11), and within complex environments like triple junctions (Figure 12(a)). The model performance is further validated by predicting $E_{den}$ for a larger nanocrystal (~25 nm$^3$) with 10 differently oriented grains (Figure 12 (b)). The combined model outperforms previously reported data-driven models predicting atomic properties for nanocrystalline materials. Figures 11 and 12 demonstrate the superior predictive capability of our model, which is due to the large variety of atomic environments in our database of metastable states and the power of SFDs in representing LAEs.

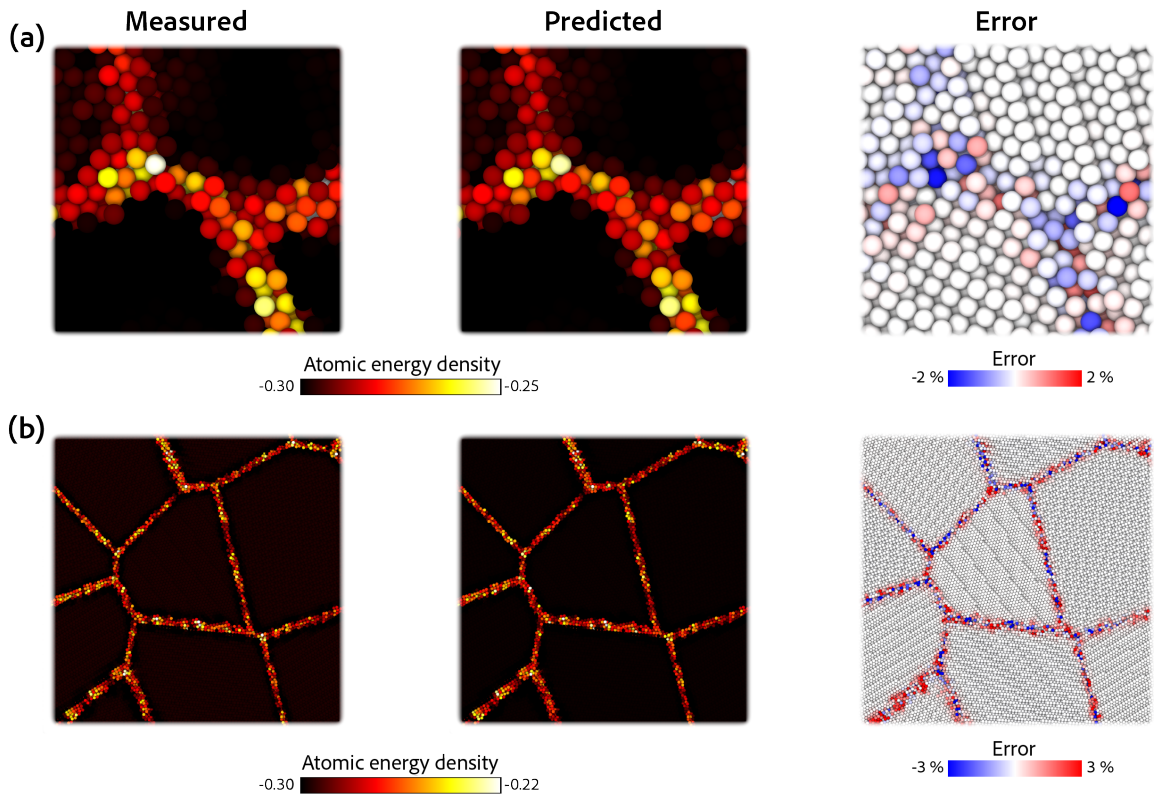

*Figure 12, Actual and predicted atomic energy density and prediction error for (a) triple junction from nanocrystal (10 nm³) with 15 grains and (b) nanocrystal (25 nm³) with 10 grains. ML model predicts the atomic energy density for nanocrystals (25 nm³ and 10 grains) with an error of ±3 %. The three-dimensional perspective view of (b) is shown in Figure S17.*

## Conclusions

In conclusion, our study effectively demonstrates the proficiency of physics-based descriptors in accurately characterizing local atomic environment at GBs for different symmetric tilt boundaries. Utilizing advanced data-driven machine learning methods, we established a robust structure-property relationship that enables the prediction of grain boundary energy in metastable structures with unprecedented accuracy. Our approach uses two-class clustering (based on GMM) method to identify the GB region, which is shown to provide a high-fidelity representation of the GB. The average SFD values as features show exceptional performance over other considered descriptors for predicting GBE. Moreover, derived features, such as GMM classes or similarity metric help to develop a fundamental understanding of the relationship between deformation modes at GB and the property of interest. We conclusively show that GBE is predominantly governed by third (strain gradients) and fourth order (2$^{nd}$ derivatives of strain) deformations at the GB. Further, our model for energy density developed using only 19 physically-motivated features (SFDs) shows superior predictive capability: model, which is trained only on symmetric tilt boundaries, predicted the energy density for unseen nanocrystals, with arbitrary GBs and triple junctions, with a maximum error of ±2%. This remarkable accuracy is attributed to comprehensive sampling of a variety of atomic environments in our database with metastable symmetric tilts and the predictive power of SFDs.

*May 31, 2024*

## Methods

**Grain boundary (GB) database**

To develop structure-property relationships, we created a GB database comprised of minimum-energy and metastable structures of <100>, <110>, <111>, and <112> symmetric tilt boundaries of FCC Cu. The interaction between Cu atoms is represented using EAM potential developed by Mishin et al.[64]. The database is created by considering macroscopic and microscopic degrees of freedom for symmetric tilt GBs. The macroscopic space defines the misorientation (rotation) between two grains and GB planes. As these are symmetric tilts, the GB planes for both grains have the same orientation with respect to the individual grain axes, and the grains are rotated by an equal and opposite angle around the rotation axis. The misorientation axis for symmetric tilt boundary is perpendicular to the GB normal and lies in the GB plane. Further, microscopic degrees of freedom are also considered which involve the translation of grain in the GB plane. Such translations are accommodated by considering the overlapping distance between atoms of both grains at the GB plane so that a good local atomic density is maintained. Inclusion of these microscopic degrees of freedom leads to the formation of a large number of metastable GB structures.

The GB rotation vector list was created from previously known minimum energy structures based on $\gamma$-surface approach[65-67]. These structures are periodic in the GB plane and non-periodic in the normal direction. To find the minimum energy and metastable structures from these initial bicrystals, one grain is translated with respect to the other in the GB plane on a grid of 4 × 4 evenly spaced points covering that periodic domain. Overlapping atoms within a specified cutoff radius are deleted to generate various structures. The cutoff radius was varied from 1.5 to 2.5 Å with a step of 0.25 Å, more details about the methodology for creating GB structures can be found in ref.[67]. The initial GB structures generated by this were then relaxed (minimized) in LAMMPS[68,69] by conjugate gradient minimization to zero pressure. This procedure would allow the structures to relax in the direction normal to the GB effectively allowing some sampling in displacements in that direction and allowing the constructed GB to optimize their local densities. The GBE for these minimized GB structures using the following equation:

$$GBE = \frac{\sum_i^n E_i - n \times E_{Cu}}{A},$$

where $E_i$ is the atomic energy of a Cu atom at the GB, $n$ is the number of atoms at GB, $E_{Cu}$ is the cohesive energy of FCC Cu, and $A$ is the GB area. As mentioned above, the structure is not periodic normal to GB; thus, GBE is calculated around the GB plane at the center of bicrystal. A total of 5479 GBs for four symmetric tilts of Cu is generated, as represented in Figure 1.

Furthermore, as GBE is an average property, the ratio of computed atomic energy ($E_i$) and Voronoi volume ($V_{vor}$) for each atom in the microstructure is also used to calculate the atomic energy density ($E_{den}$). The atomic energy density ($E_{den}$) is calculated by using following equation:

$$E_{den} = \frac{E_i}{V_{vor}}$$

The nanocrystalline Cu microstructure with different grain sizes is generated using Voronoi tessellation methods as implemented in Atomsk[70]. These microstructures are minimized to zero pressure using Isothermal–isobaric ensemble (NPT). We calculated the atomic energy density for nanocrystals to apply the developed ML model on untrained data.

**Strain Functional Descriptors (SFD)**

To identify the LAE at a GB, we used the recently developed SFDs[60]. The SFDs are generated from a Taylor series expansion of the local number density at each atom and thereby carry information about the local derivatives of that density. A Gaussian function with a finite cutoff radius at the numerical noise level (e.g. where the Gaussian weighting is ~$10^{-6}$) is used to map the atomic number density onto a continuum field. The SFD basis is developed analogously to the principal quantum number labeling of atomic orbitals which tracks their behavior in terms of angular momentum eigenvalues. This makes SFDs physically informed with regard to the underlying nature of the electronic structures and how those would adjust under these changes in strains. The local moment tensors, resulting from the Taylor series expansion of the number density, are mapped to rotationally-invariant descriptors via the Clebsh-Gordan coupling procedure. This describes the local neighborhood distortions in a manner independent of the external frame. The SFDs are represented as PX_IY, where X is tensor order (corresponding to the order in the Taylor series expansion), and Y is the internal ranking within that order. The prefix "P" specifically indicates that these descriptors map the positions of atoms within the structure. Additional descriptors can be developed for the net orientation of the object with respect to some ideal frame, vector quantities (such as displacement, force, velocity) and higher-order tensors. A comprehensive description of the development of SFDs can be found in the article by Kober et al.[60].

Here, a reduced set of 19 SFDs, encompassing shape and size descriptors but not internal orientation functions, are considered to characterize the LAEs. SFDs do not require information about the initial or reference structure (as the reference state is a spherically symmetric distribution of number density) and any changes in the atomic arrangement is effortlessly captured. For example, SFD vectors such as P0_I0, P2_I2, and P4_I8 represent the density terms (with different weightings with respect to the radial distance) and correlate to excess volume at the GB, whereas P2_I0 and P2_I1 are the net shear/deviatoric strain (similar to von Mises strain invariant) and character of the shear (orthorhombic vs. tetragonal distortion) respectively. Strain gradient terms, such as P3_I0 and P3_I4, are zero for the ideal FCC environment. Hence, the amount and nature of the deviation of GB atoms from the perfect crystal structure is readily quantified. Sigma (width of the Gaussian) and radius (cutoff) parameters used for generating SFDs for FCC Cu are 0.909 and 5.904 Å, respectively. Guidelines for choosing optimal values for these parameters can be found in Kober et al.[60].

**Machine learning**

To develop ML models for establishing structure-property relationships, we start with generating SFDs for GB structures. Unlike previous efforts in which GB regions are identified based on traditional structure characterization methods with a fixed cutoff, we first develop a two-class unsupervised Gaussian Mixture Model (GMM)[61] using per-atom SFDs to identify the GB region. GMM is a probabilistic model that considers the distribution of a data point in the dataset is generated from a mix of various Gaussian distributions with unknown parameters. GMM employs the expectation-maximization (EM) algorithm to calculate parameters for mixture models by using maximum likelihood estimation. GMM model effortlessly isolates the GB region at the center and outer edges (normal to GB) of the GB. Compared to disordered atoms at GB, characterized using common neighbor analysis or polyhedral template matching, two-class GMM shows that the GB region extends farther away (Figure S1), thus enhancing the fidelity of characterization. The central region, as shown in Figure S1, is referred to as the "GB region" and is considered for developing ML models. Section S1 in supporting information discusses the ability of SFDs to characterize the GB region.

We adopted two approaches for developing regression models using SFDs. The first approach consists of two possible methods. In the first method, we use statistical values (mean and kurtosis) of GB atom SFDs as features in ML models. For example, averaging SFDs of all the atoms in the GB region results in the representation of a GB by one mean SFD vector. This reduces the feature space for a given GB from the number of atoms × 19 SFDs to only one SFD vector with 19 components (Figure 2(b)-i). The second method uses SFDs for every atom in the GB as features in ML models (Figure 2(b)-ii). This is used to train the regression model for atomic energy density prediction.

Features for the second approach are derived using GMM or similarity matrices. First, we develop a six-class GMM model using SFDs for all GB atoms. Each GMM class represents distinct LAEs. The relative importance of these classes in describing a particular GB structure enables us to derive crucial physical insights using SFDs. The first method in this approach uses a set of features based on the probability of each atom belonging to a specific GMM class. The average values of these probabilities are considered as representative features, as shown in Figure 2(b)-iii. The second method is based on the frequency of different GMM classes in each GB structure. GMM assigns each atom at the GB to a particular class; thus, each GB is comprised of atoms belonging to six different classes. Therefore, we can measure the number of atoms belonging to a particular class in a GB and use the normalized frequency (i.e., divided by the total number of atoms at GB) of different classes as features. Similar to mean SFD features, each GB is represented by a vector of six components (Figure 2(b)-iv). The final set of features for the second approach are derived by measuring the cosine similarity ($similarity(\vec{a}, \vec{b}) = \frac{\vec{a}.\vec{b}}{|\vec{a}||\vec{b}|}$) between the SFD vectors describing the GMM class corresponding to FCC atoms ($\vec{a}$) and rest of the GB atoms ($\vec{b}$). Atoms having higher cosine similarity (~1) are close to the FCC structure in that symmetric tilt boundary, whereas lower values represent highly deformed atoms. Next, the cosine similarity value is discretized to different classes, and the frequency of respective classes is used as a feature for model development (Figure 2(b)-v). The importance and physics of different feature sets in predicting properties are discussed in respective sections. A complete schematic depicting the various models is shown in Figure 2. The Scikit-learn[71] python package is used for developing Unsupervised GMM and Random-Forest regression models.

**Data availability**
The data that support the findings of this study are available from the corresponding authors upon reasonable request.

**Code availability**
All codes needed to evaluate the conclusions in the paper are available from the corresponding authors upon reasonable request.

**Acknowledgment**
Simulations were performed using the HPC at Los Alamos National Laboratory (LANL). The authors gratefully acknowledge support from the U.S. Department of Energy through the LANL/LDRD Program for this work. The authors acknowledge funding from the LDRD-DR "Investigating How Material's Interfaces and Dislocations Affect Strength" (iMIDAS) under grant no. 20210036DR [Abigail Hunter, PI and Saryu J. Fensin, Co-PI]. A.M., N.M., and E.M.K. also acknowledge funding from the LDRD project 20220814PRD4: "Grain Boundary

**Author contribution**
A.M., N.M., and E.M.K. designed and performed the study. S.A.S. and S.J.F. generated the GB structure for the database and calculated GBE, and A.M. and N.M. calculated the atomic energy density. All authors discussed and formulated the results. A.M. wrote the first draft of the manuscript, with S.J.F., N.M., and E.M.K. finalizing it. E.M.K. developed the initial SFDs, and A.M. and N.M. generated the collective descriptors and developed the machine learning models. S.J.F did the funding acquisition.

**Declaration of Competing Interest**
The authors declare that they have no known competing financial interests or personal relationships that could have appeared to influence the work reported in this paper.

*Supplementary Information for*
# Learning from metastable grain boundaries

**Avanish Mishra[1], Sumit A. Suresh[2], Saryu J. Fensin[2], Nithin Mathew[1*], and Edward M. Kober[1*]**

[1]Theoretical Division (T-1), Los Alamos National Laboratory, Los Alamos, 87545, NM, USA

[2]Materials Physics and Applications (MPA-CINT), Los Alamos National Laboratory, Los Alamos, 87545, NM, USA

*mathewni@lanl.gov (Nithin Mathew) and emk@lanl.gov (Edward M. Kober)

### Section S1: Grain boundary region

Figure S1 shows GB structures characterized using traditional common neighbor analysis (CNA) and a two-class unsupervised Gaussian mixture model (GMM) developed using SFDs. These are minimum and metastable GBs of Cu with <112> rotation axis. The GB region characterized by CNA in minimum energy GB (Figure S1(a)) is three atomic-layers thick, whereas GMM shows it extends to two more atomic layers in both grains. Similarly, metastable GB with GBE of 409 mJ/m$^2$ has some deleted overlapping atoms at the interface. However, CNA shows a similar GB region as the minimum energy structure. In contrast, the GMM model readily shows the modification in the GB region due to atom deletion, as shown in Figure S1(b). Such modifications are also observed for other metastable GBs, as shown in Figures S1(c) and S1(d).

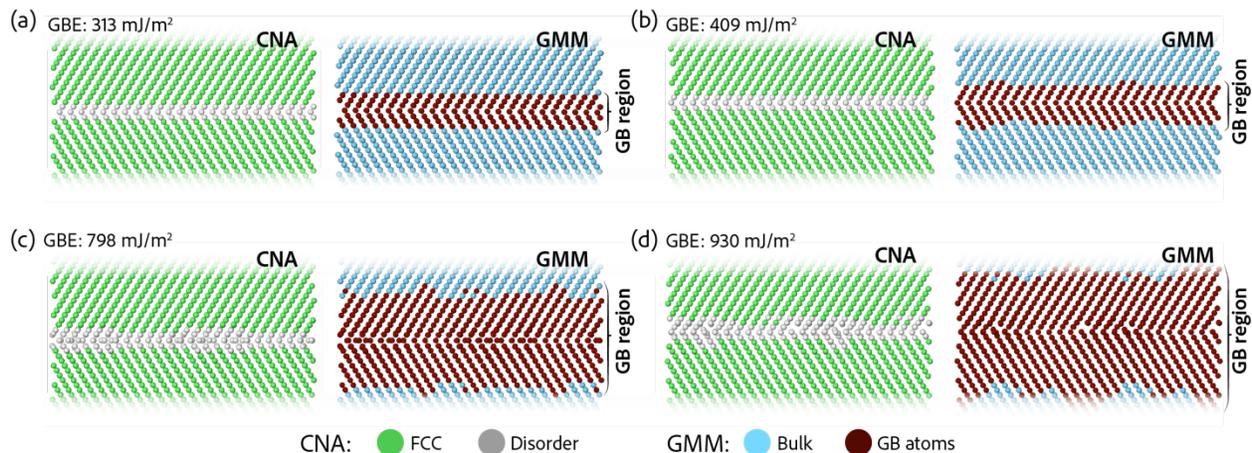

*Figure S1, Snapshots of <112> symmetric tilt GBs colored using common neighbor analysis (CNA) and two-class GMM model for Σ11GBs with GBE of (a) 313 mJ/m$^2$, (b) 409 mJ/m$^2$, (c) 798 mJ/m$^2$, and (d) 930 mJ/m$^2$.*

### Section S2: Machine learning model for GBE prediction

We tested the robustness of the developed ML model by different splits of training and validation data for model training, as shown in Figure S2. The best model will have the highest R$^2$ value and will lie at the most right of the distribution, whereas the peak in these plots indicate average performance for the developed model.

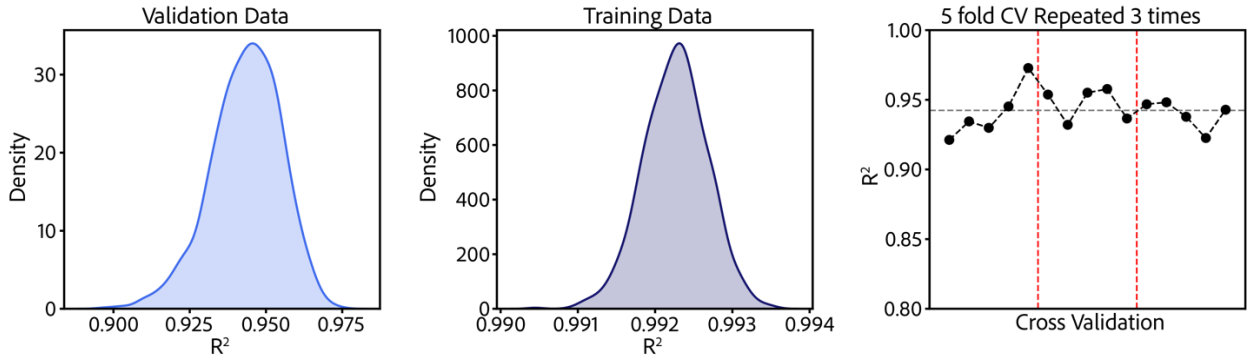

*Figure S2, Different train-validation split GBE prediction coefficient of determination ($R^2$) histogram for (a) validation and (b) training data of <112> symmetric tilt boundaries. (c) Training $R^2$ for 5 fold cross validation repeated 3 times for <112> symmetric tilt boundaries.*

Figure S3 shows the performance of GBE prediction ML models developed using only the three most important features identified in Figure 4 of the main text. For <100> GBs, the top three features are P3_I0, P4_I6, and P4_I8; for <111> GBs, they are P3_I1, P4_I6, and P4_I8; for <110> GBs, the features are P2_I0, P4_I6, and P4_I8; and for <112> GBs, the leading features are P2_I0, P4_I7, and P4_I8. Although these models exhibit a decline in performance relative to the model developed using 19 SFDs, their prediction accuracy remains exceptionally high.

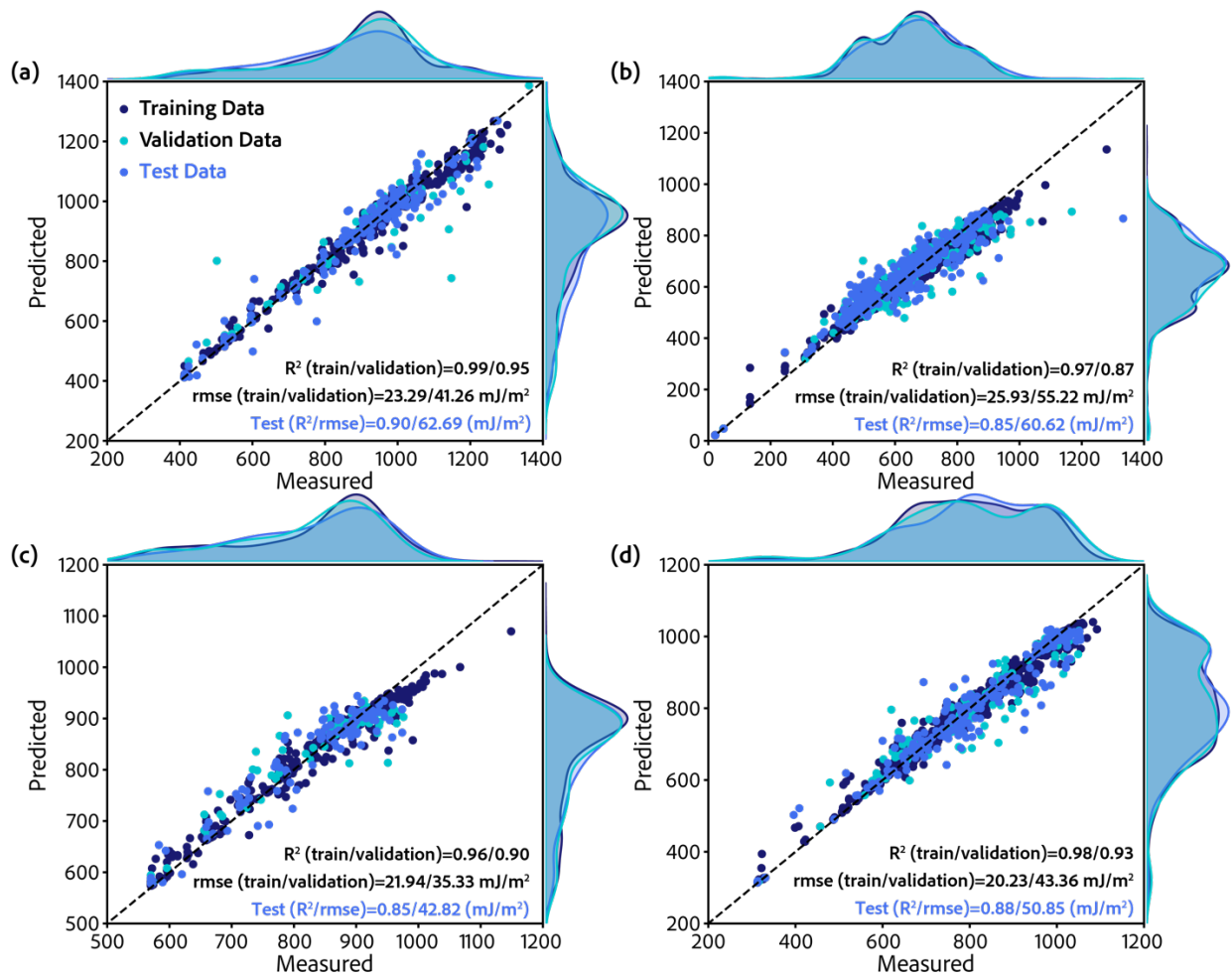

*Figure S3, Scatter plot for GBE showing measured vs predicted values for (a) <100>, (b) <110>, (c) <111>, and (d) <112> symmetric tilt boundaries for model developed using only three most important features from Figure 4 of the main text for respective tilt boundaries. Here, test is performed on 10 % of data that the model has not seen during training.*

*May 31, 2024*

## Section S3: Bivariate distributions

Bivariate distributions for mean SFDs of <100>, <110>, and <111> symmetric tilt GBs are shown in Figures S4(a), S4(b), and S4(c), respectively.

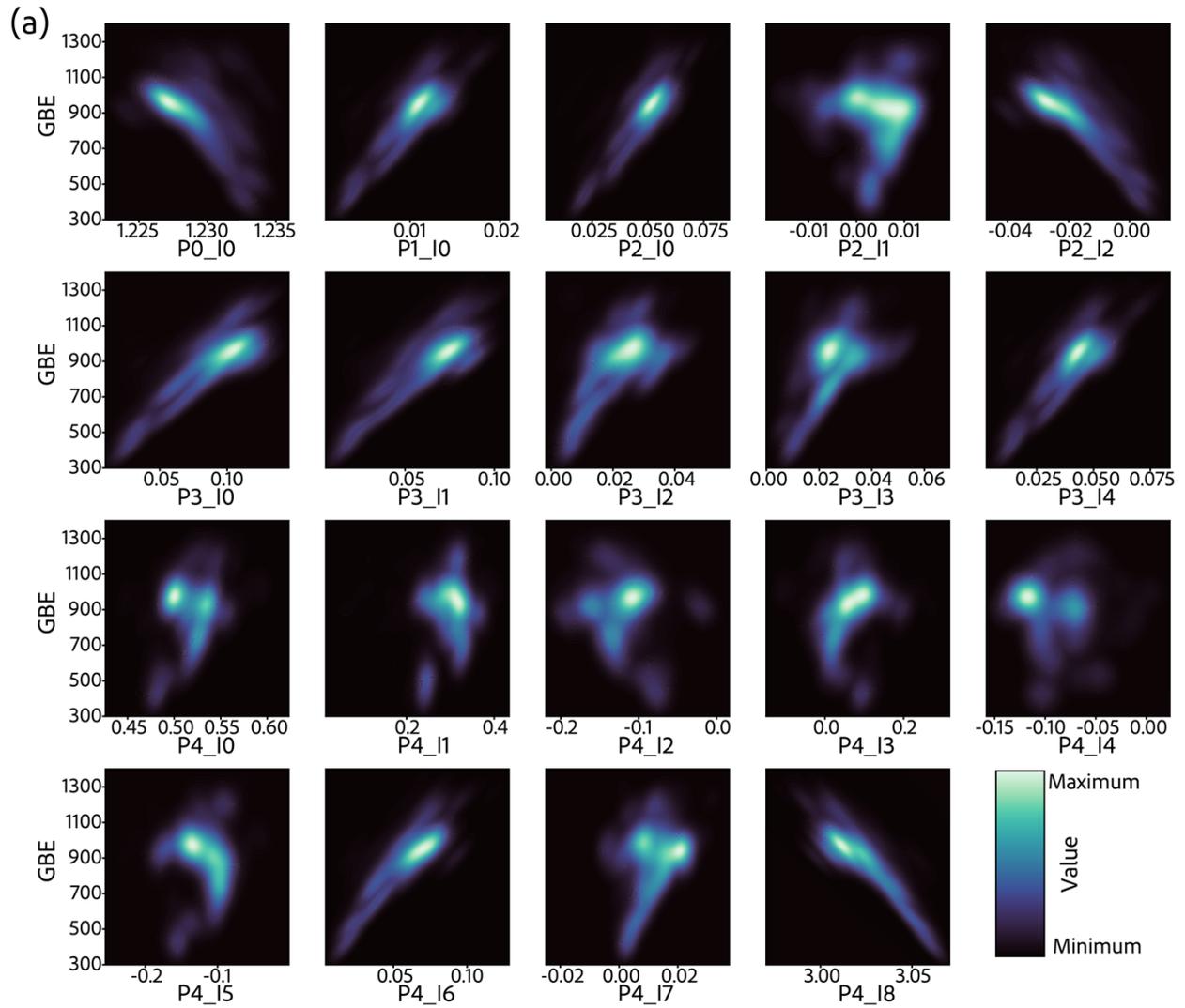

(a)

(b)

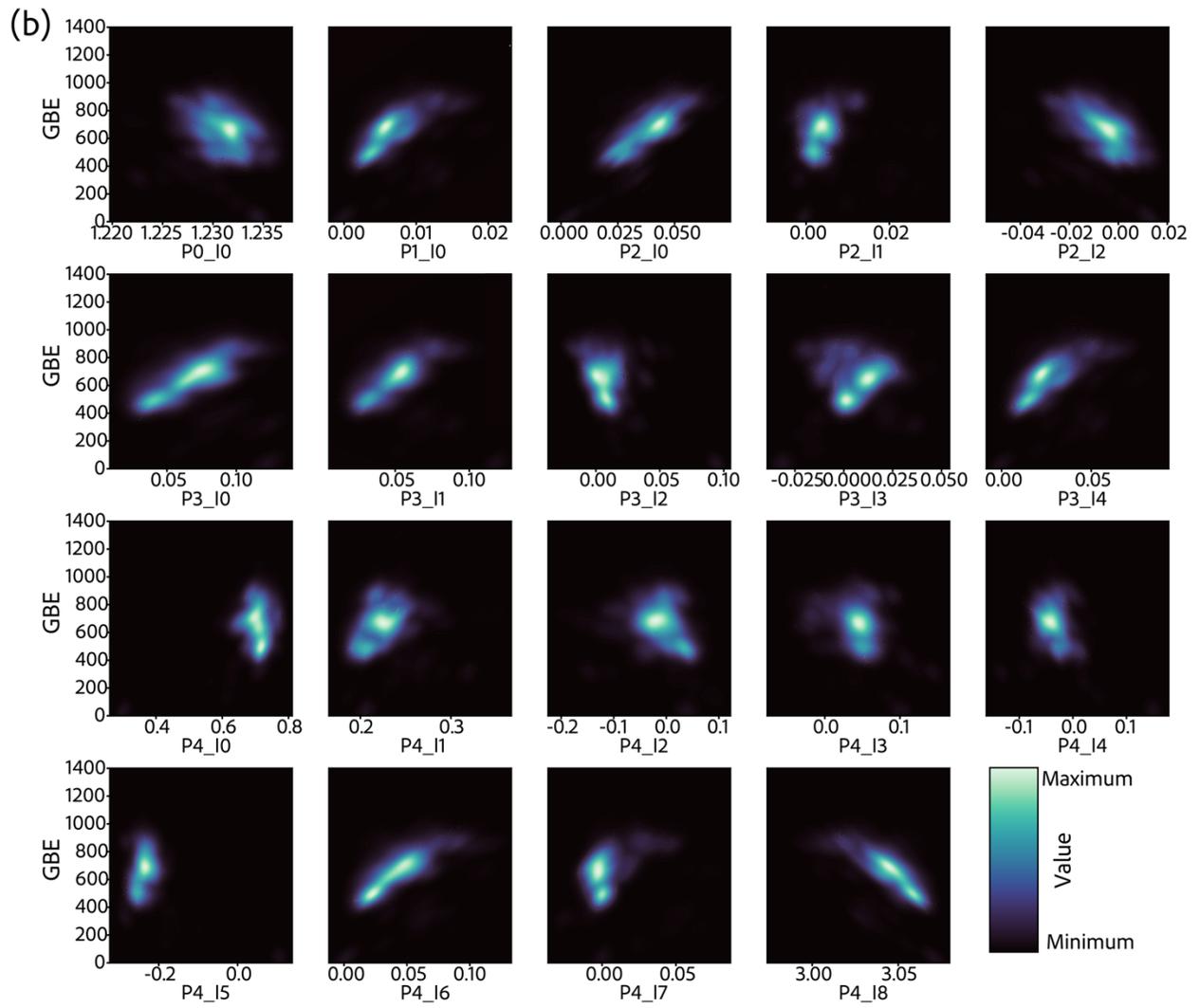

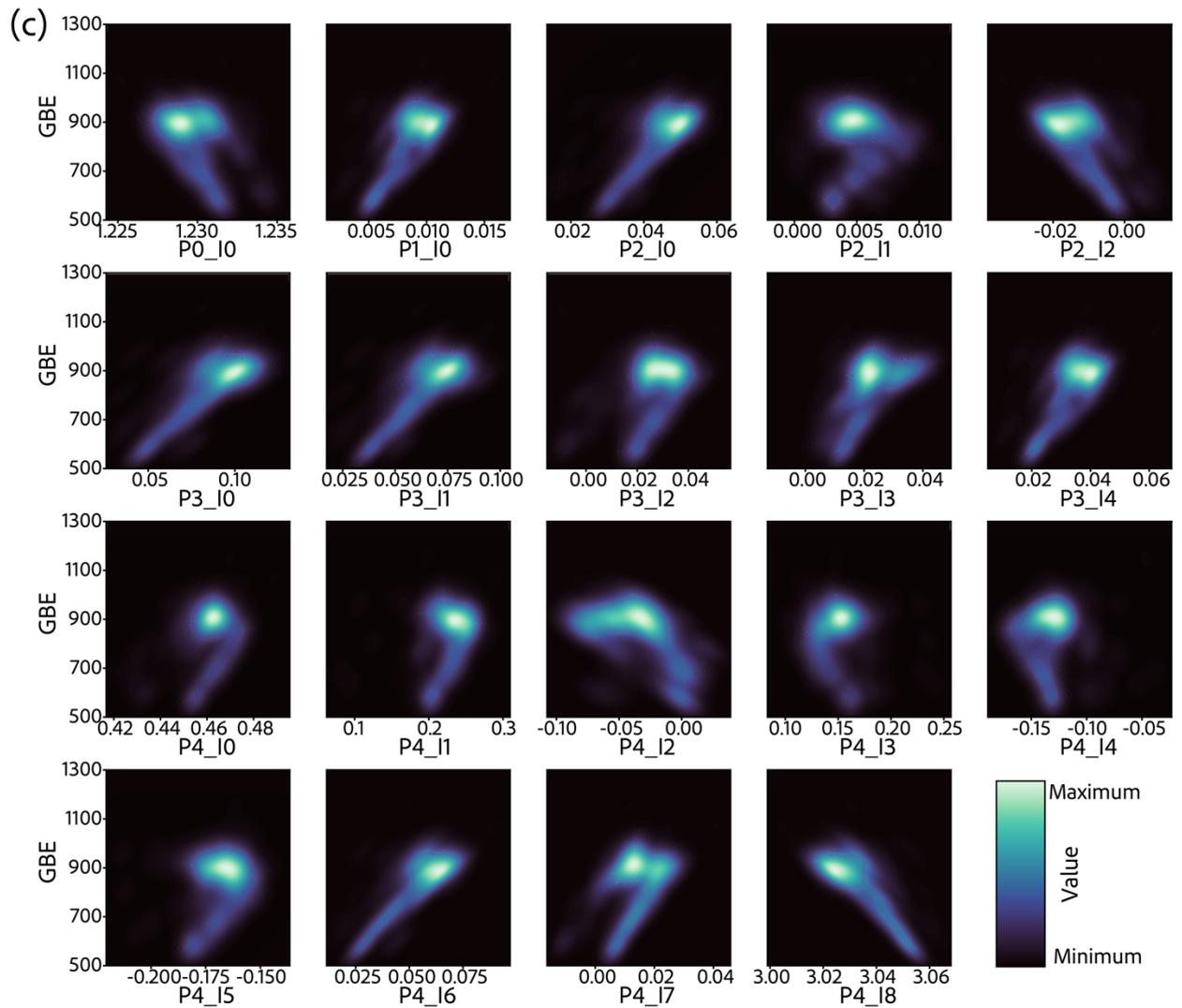

Figure S4, Bivariate distributions of grain boundary energy (GBE) and SFDs using kernel density estimation for (a) <110>, (b) <110>, and (c) <111> symmetric tilt boundaries, respectively. Features of higher importance show linear variation with GBE.

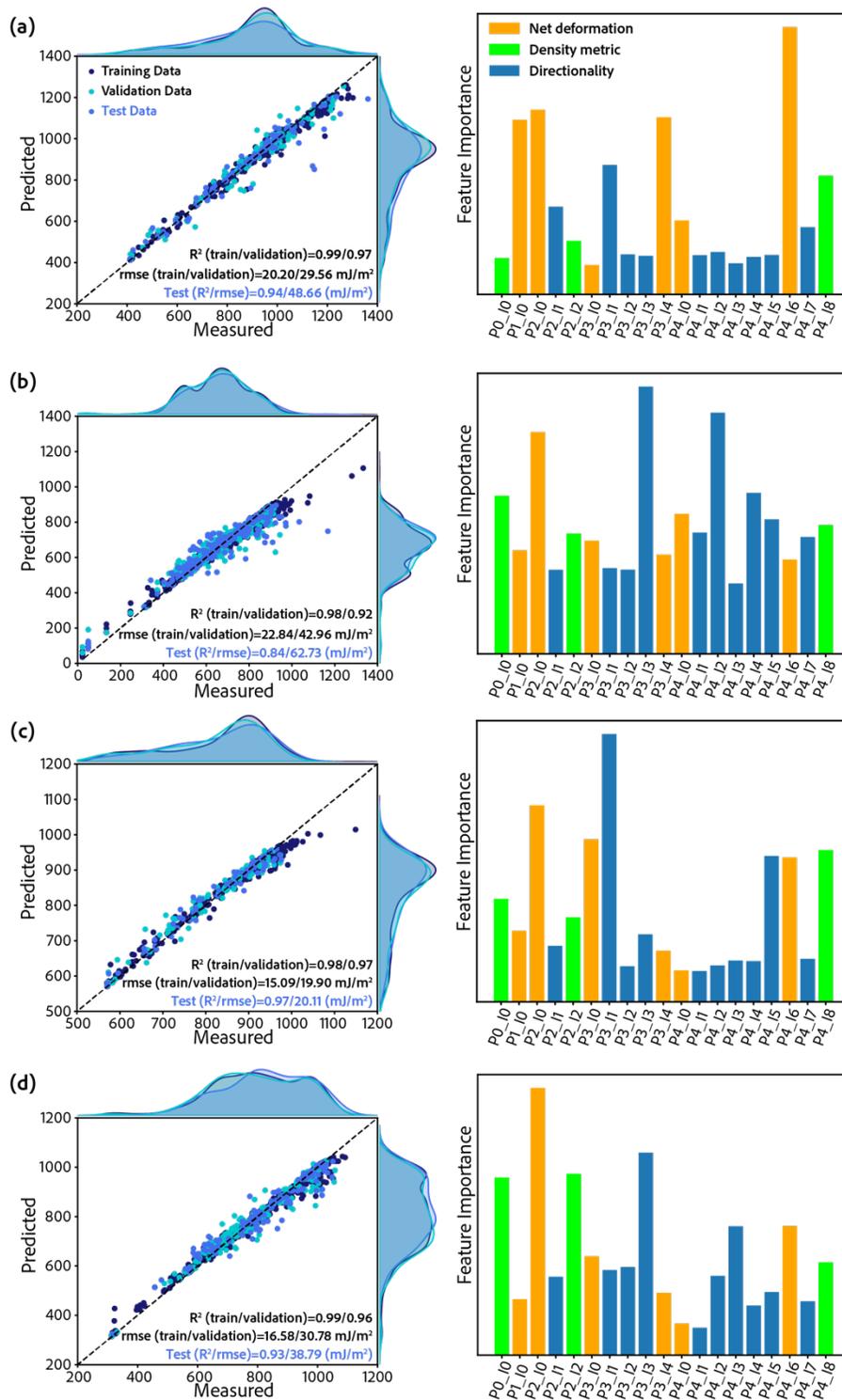

Figure S5, Scatter plot for GBE, showing measured vs predicted values and feature importance for (a) <100>, (b) <110>, (c) <111>, and (d) <112> symmetric tilt boundaries for model developed using Kurtosis of SFDs. The value of feature importance is arbitrary and the height of tower in respective plots shows important features. Net deformation, density metric, and directionality terms of SFDs are shown by orange, green, and blue bars.

*May 31, 2024*

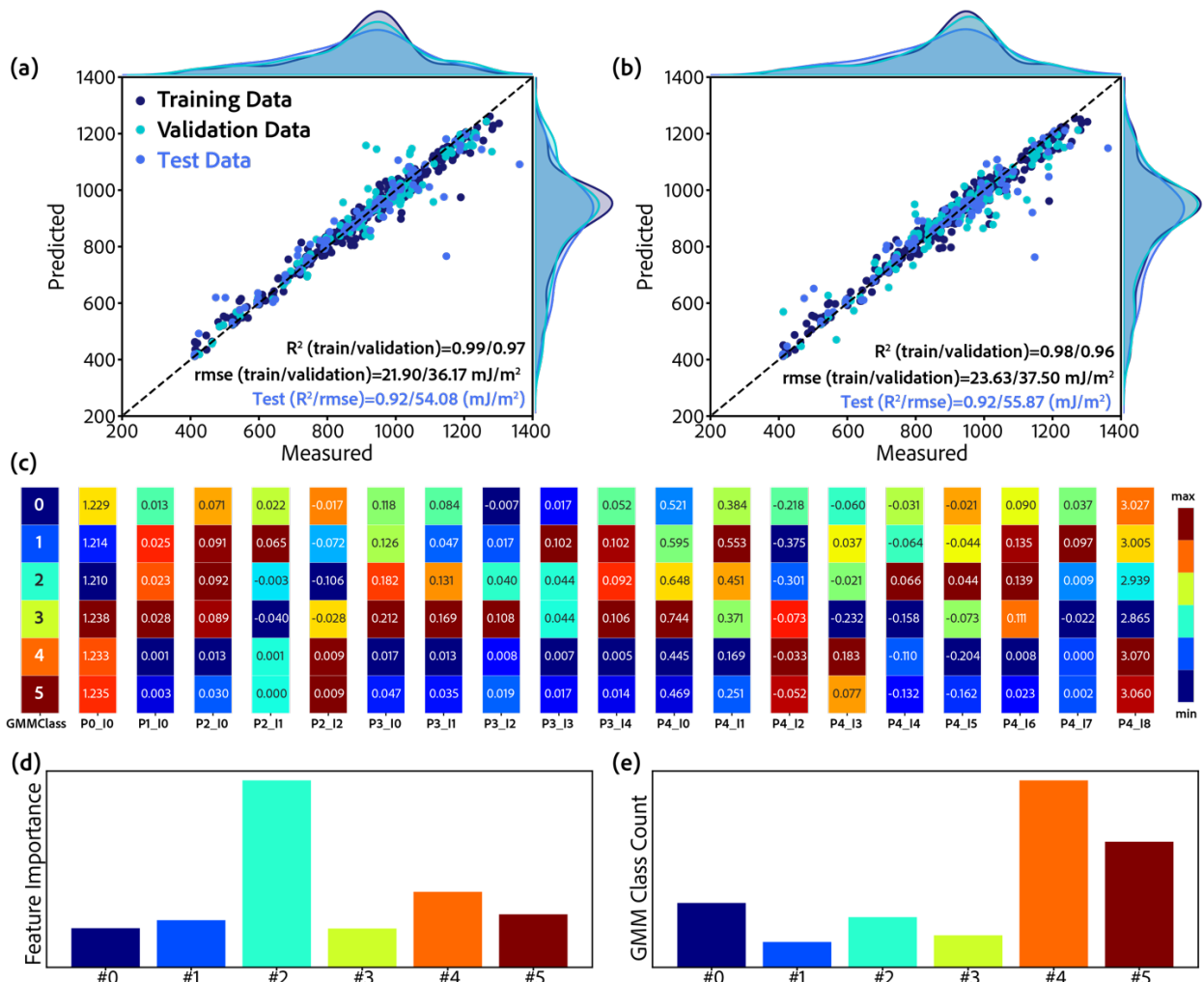

Figure S6, Scatter plot showing measured vs. predicted values for model developed using (a) average GMM class probability and (b) frequency of GMM classes as features for <100> symmetric tilt. (c) Mean value of SFDs for each GMM classes associated with the <100> symmetric tilt. Each column is colored using "jet" color map, blue-to-red. (d) Feature importance for the model developed using frequency of GMM classes as features and (e) count of different GMM classes in <100> symmetric tilt boundaries.

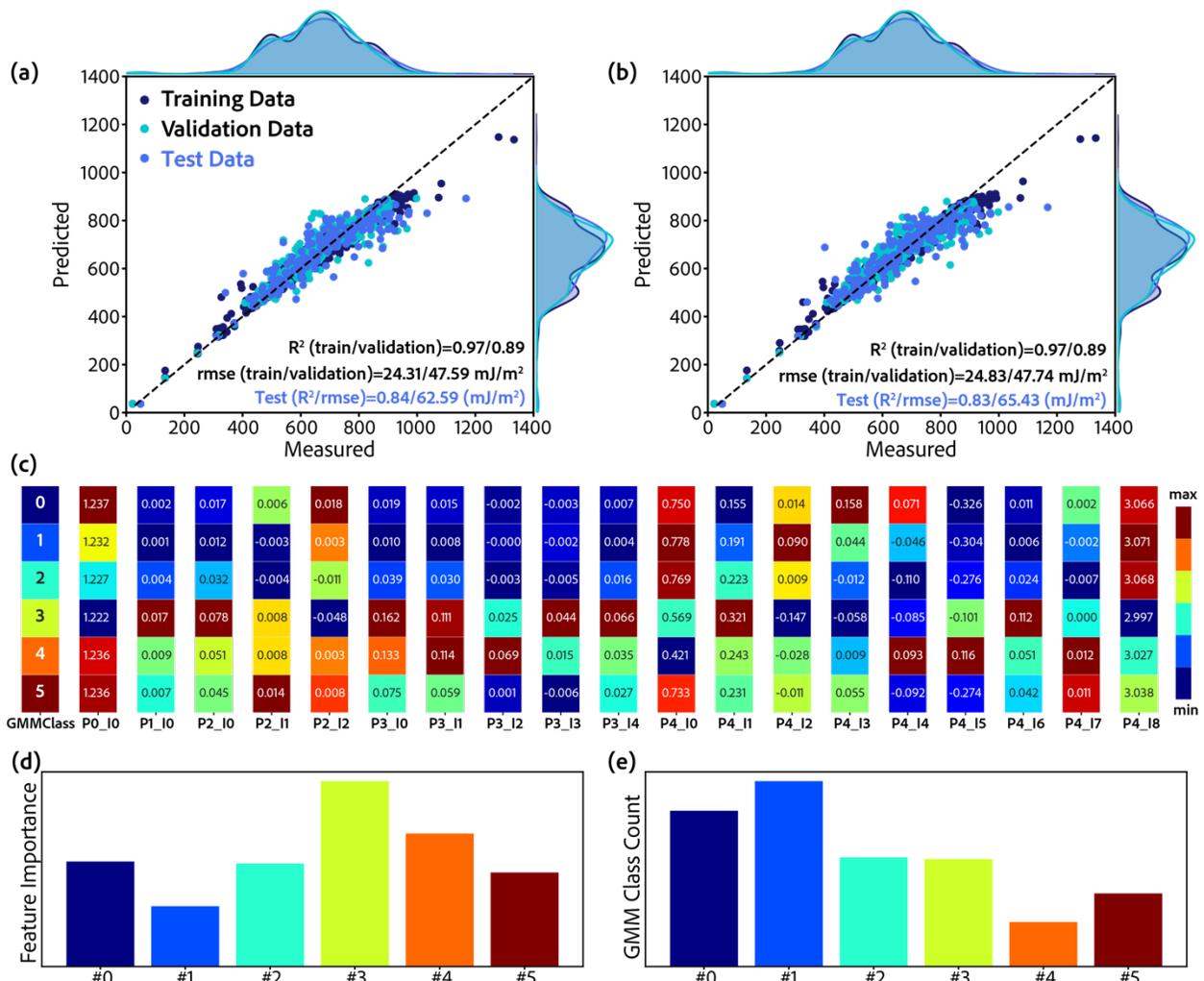

Figure S7, Scatter plot showing measured vs. predicted values for model developed using (a) average GMM class probability and (b) frequency of GMM classes as features for <110> symmetric tilt. (c) Mean value of SFDs for each GMM classes associated with the <110> symmetric tilt. Each column is colored using "jet" color map, blue-to-red. (d) Feature importance for the model developed using frequency of GMM classes as features and (e) count of different GMM classes in <110> symmetric tilt boundaries.

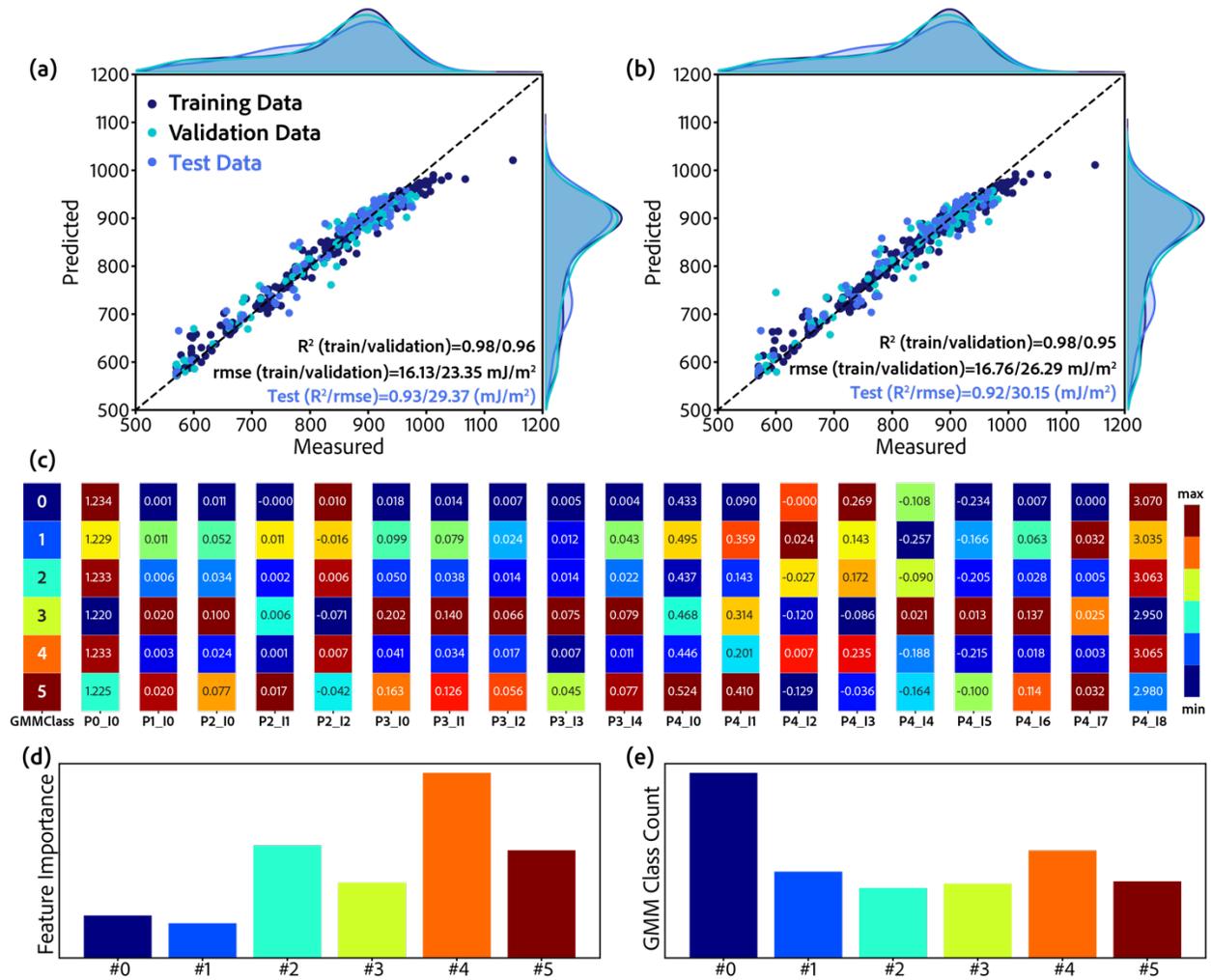

*Figure S8, Scatter plot showing measured vs. predicted values for model developed using (a) average GMM class probability and (b) frequency of GMM classes as features for <111> symmetric tilt. (c) Mean value of SFDs for each GMM classes associated with the <111> symmetric tilt. Each column is colored using "jet" color map, blue-to-red. (d) Feature importance for the model developed using frequency of GMM classes as features and (e) count of different GMM classes in <111> symmetric tilt boundaries.*

**Section S4: Regression model using cosine similarity**

The cosine similarity is a continuous variable, unlike GMM classes. However, to build an ML model and get insights using similarity, we discretized the cosine similarity metric using ten equally-spaced bins (0-9). Note that the binning is not done based on distribution but range. We tested similarity features for <112> symmetric tilts; the ML model has $R^2$ of 0.92 for validation data, which is less than that of ML models with GMM probability/frequency features. Although, similar to those models, the feature importance shows that lower similarity (bin #6 and #7) features are most important as shown in Figure S9. In summary, we trained various ML models using different features for individual symmetric tilt boundaries. Among all, the ML model developed using the mean SFDs exhibits the best performance for establishing structure-property relationship for GBE. Besides, unlike descriptors used in previous studies, the application of physics-informed descriptors (SFDs) unravels the role of deformation modes for prediction of

*May 31, 2024*

GBE.

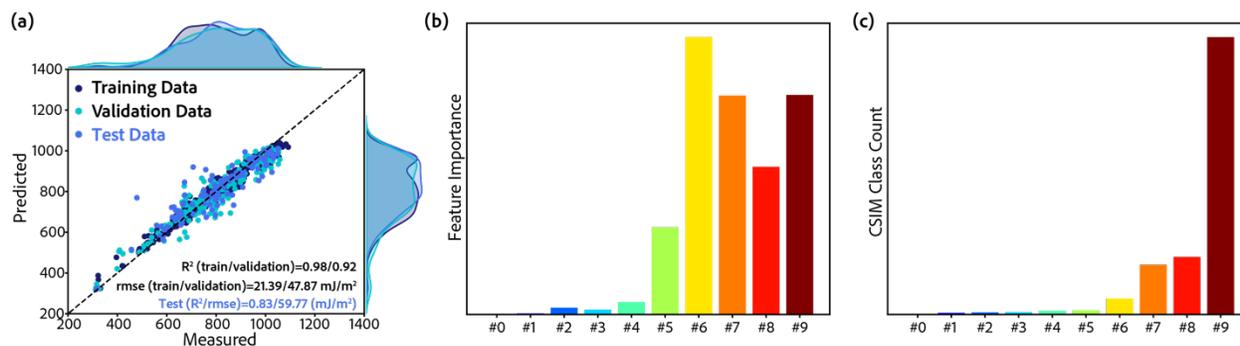

*Figure S9, (a) Scatter plot showing measured vs predicted values, (b) feature importance, and (c) count of different cosine similarity classes in <112> symmetric tilt boundaries for model developed using discretized cosine similarity as features. The value of feature importance is arbitrary and the height of tower in respective plots shows important features.*

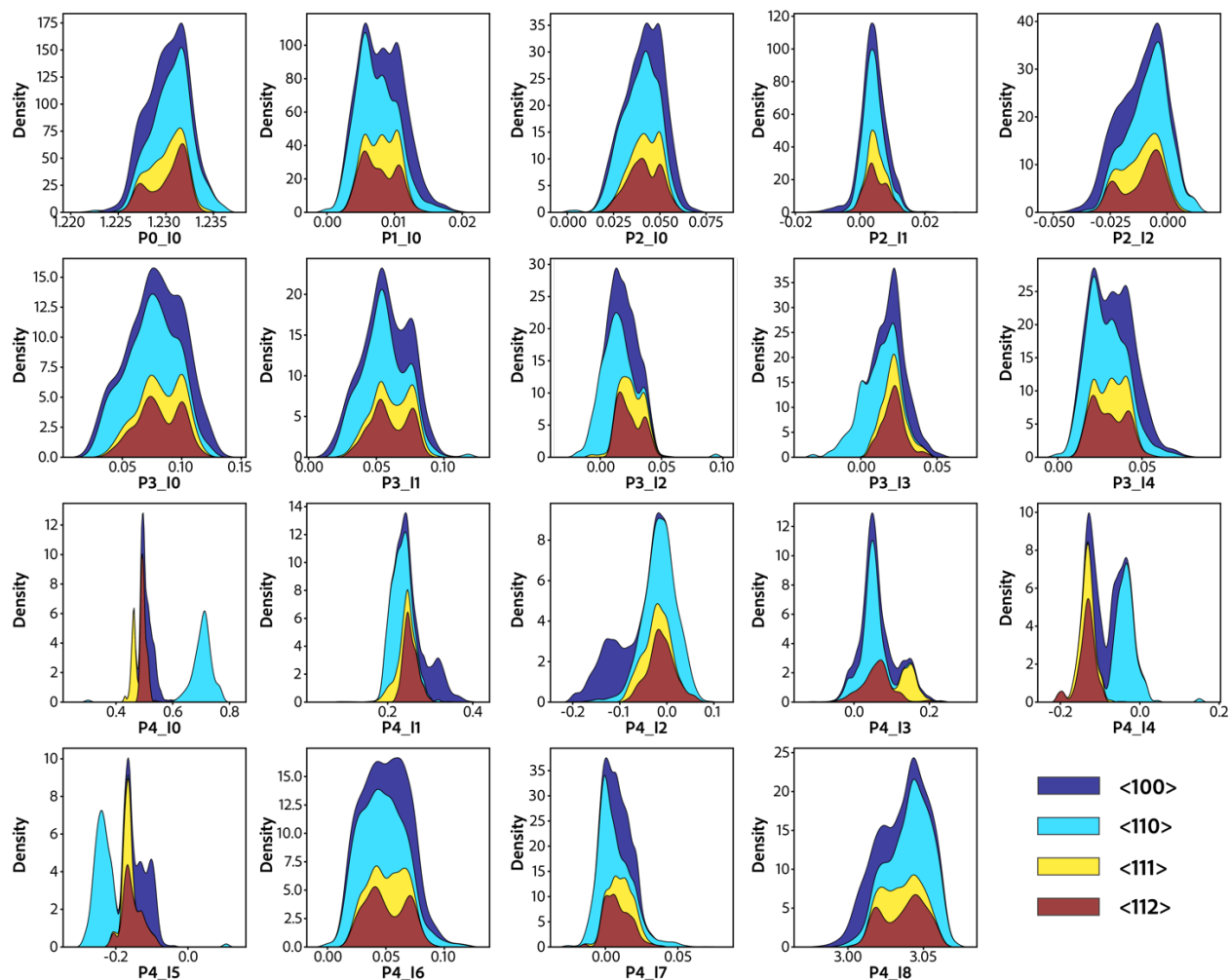

*Figure S10, Distribution of 19 mean SFD features for different symmetric tilt boundaries showing variation in the SFD values for GB atoms.*

## Section S5: Machine learning model for atomic energy density prediction

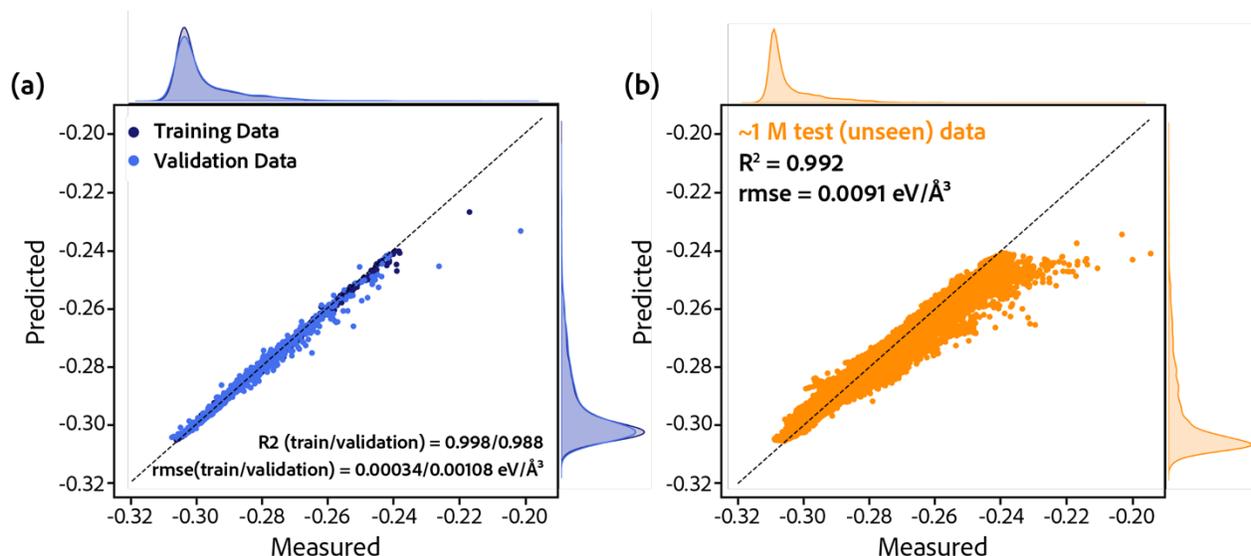

Figure S11, Scatter plot showing measured vs. predicted atomic energy density for model developed using (a) 12K datapoints for training and validation, and (b) prediction for ~1 M unseen datapoints (atoms) of <112> symmetric tilt boundaries.

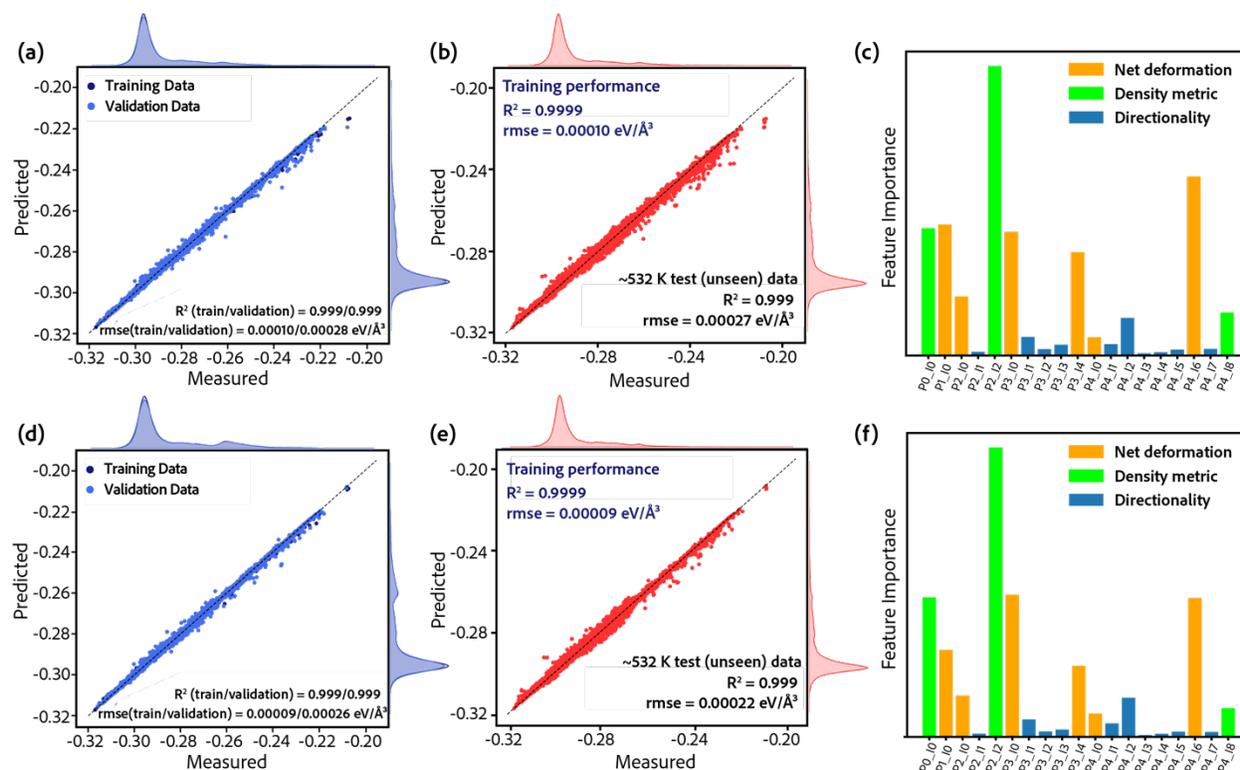

Figure S12, Scatter plot showing measured vs. predicted atomic energy density for a model trained on <100> symmetric tilt GBs dataset by (a) randomly selecting 200K data and (d) curated 200K data with the fixed outliers. Scatter plots in (b) and (e) showing measured vs. predicted atomic energy density for the rest (unseen) ~532 K data (atoms) using models developed in (a) and (d), respectively. Here data points in (a)

*are selected randomly from $E_{den}$ distribution and in (b) curated data is selected by fixing the problem with outliers; for a dataset with 200K points, 180K are randomly selected, whereas 40K are selected from the tail of the distribution, out of which 20K (randomly selected from 40K) are used to construct a 200K dataset. Feature importance for model developed in (a) is shown in (c) and for model developed in (d) is shown in (f).*

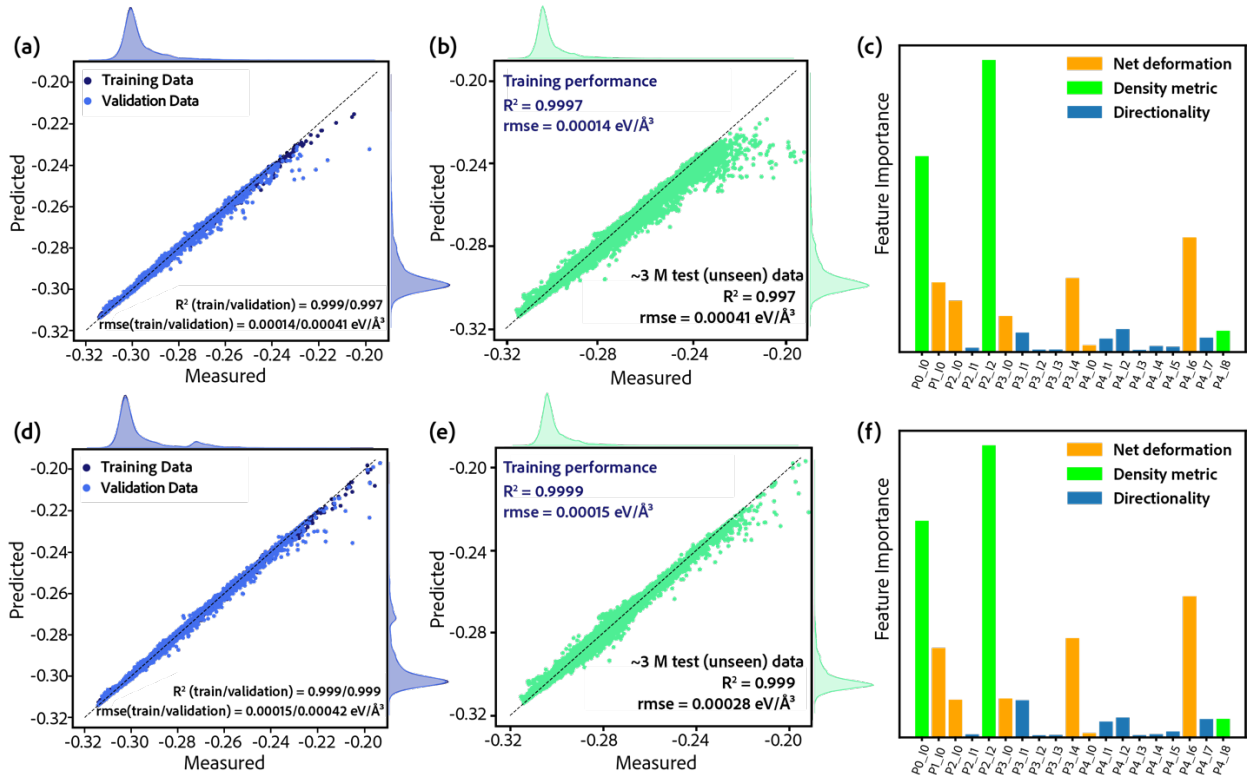

Figure S13, Scatter plot showing measured vs. predicted atomic energy density for a model trained on <110> symmetric tilt GBs dataset by (a) randomly selecting 200K data and (d) curated 200K data with the fixed outliers. Scatter plots in (b) and (e) showing measured vs. predicted atomic energy density for the rest (unseen) ~3 M data (atoms) using models developed in (a) and (d), respectively. Here data points in (a) are selected randomly from $E_{den}$ distribution and in (b) curated data is selected by fixing the problem with outliers; for a dataset with 200K points, 180K are randomly selected, whereas 40K are selected from the tail of the distribution, out of which 20K (randomly selected from 40K) are used to construct a 200K dataset. Feature importance for model developed in (a) is shown in (c) and for model developed in (d) is shown in (f).

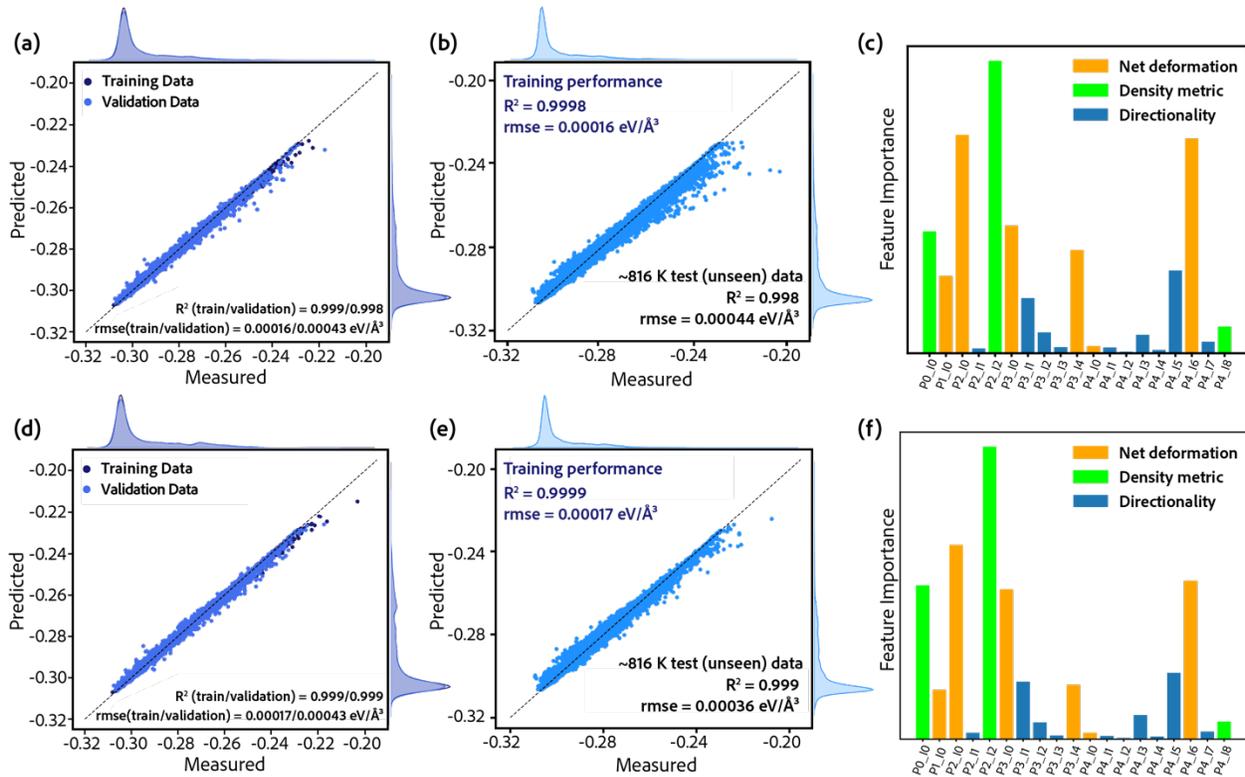

*Figure S14, Scatter plot showing measured vs. predicted atomic energy density for a model trained on <111> symmetric tilt GBs dataset by (a) randomly selecting 200K data and (d) curated 200K data with the fixed outliers. Scatter plots in (b) and (e) showing measured vs. predicted atomic energy density for the rest (unseen) ~816 K data (atoms) using models developed in (a) and (d), respectively. Here data points in (a) are selected randomly from $E_{den}$ distribution and in (b) curated data is selected by fixing the problem with outliers; for a dataset with 200K points, 180K are randomly selected, whereas 40K are selected from the tail of the distribution, out of which 20K (randomly selected from 40K) are used to construct a 200K dataset. Feature importance for model developed in (a) is shown in (c) and for model developed in (d) is shown in (f).*

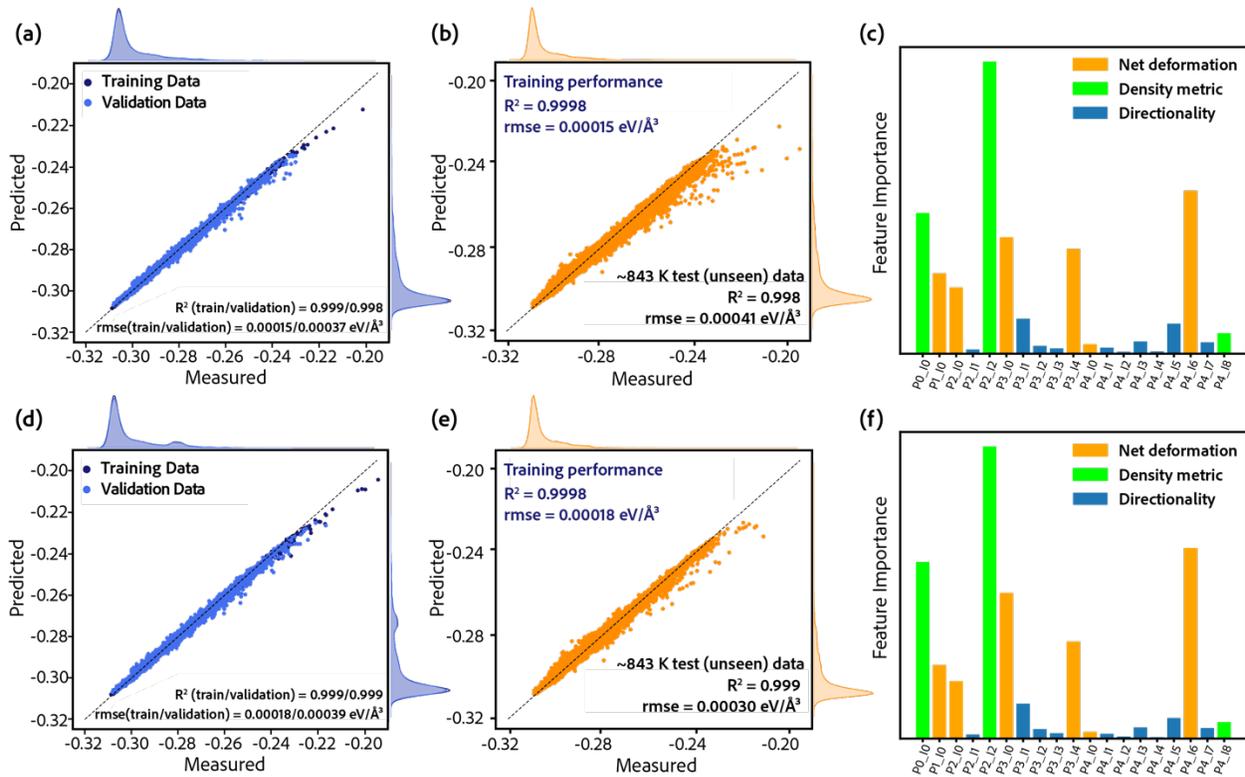

*Figure S15, Scatter plot showing measured vs. predicted atomic energy density for a model trained on <112> symmetric tilt GBs dataset by (a) randomly selecting 200K data and (d) curated 200K data with the fixed outliers. Scatter plots in (b) and (e) showing measured vs. predicted atomic energy density for the rest (unseen) ~843 K data (atoms) using models developed in (a) and (d), respectively. Here data points in (a) are selected randomly from $E_{den}$ distribution and in (b) curated data is selected by fixing the problem with outliers; for a dataset with 200K points, 180K are randomly selected, whereas 40K are selected from the tail of the distribution, out of which 20K (randomly selected from 40K) are used to construct a 200K dataset. Feature importance for model developed in (a) is shown in (c) and for model developed in (d) is shown in (f).*

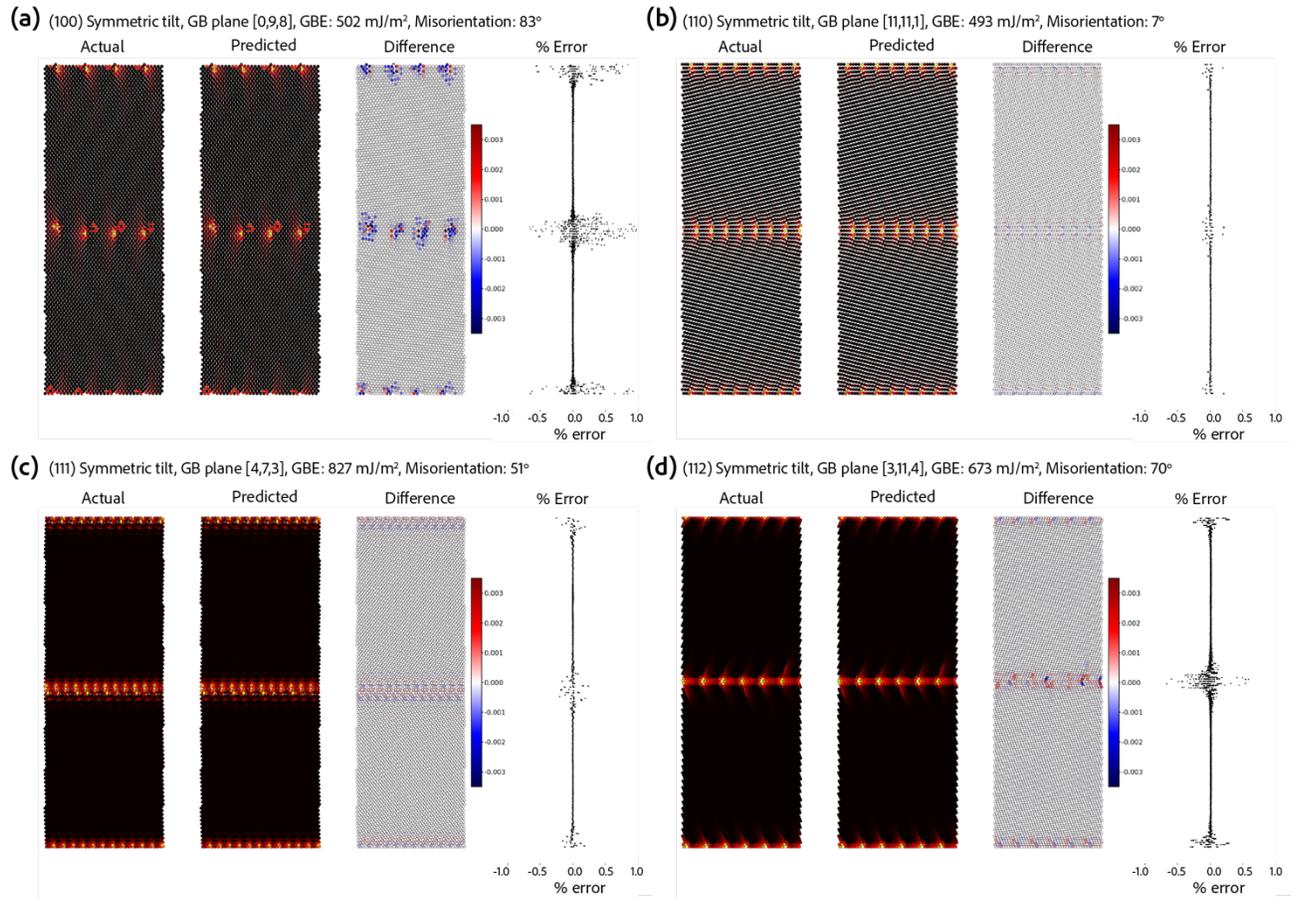

Figure S16, Actual and predicted atomic energy density and error in prediction for (a) <100>, (b) <110>, (c) <111>, and (d) <112> symmetric tilt boundaries. ML model predicts the atomic energy density for bicrystals with a maximum error of ±1 %

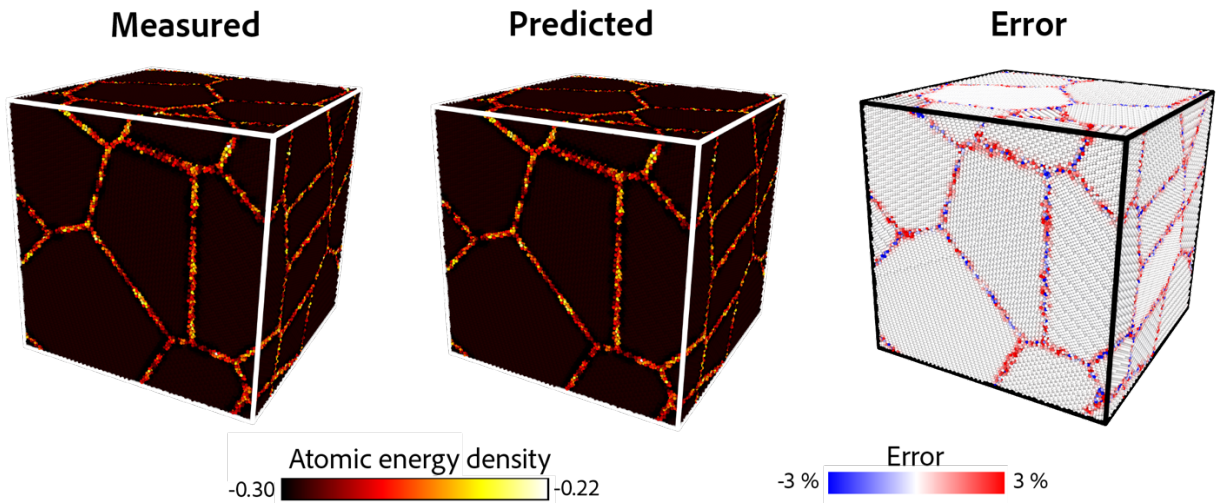

Figure S17, Actual and predicted atomic energy density and prediction error for (b) nanocrystal (25 nm$^3$) with 10 grains. ML model predicts the atomic energy density for nanocrystals (25 nm$^3$ and 10 grains) with an error of ±3 %.

*May 31, 2024*